\documentclass[a4paper,fleqn]{cas-dc}
\usepackage{natbib}

\usepackage{amsmath}
\usepackage{graphicx}
\usepackage{subcaption}

\usepackage{boxedminipage}
\usepackage{textpos}

\usepackage{url}

\begin{document} \sloppy
\let\WriteBookmarks\relax
\def\floatpagepagefraction{1}
\def\textpagefraction{.001}
\shorttitle{3D city models for urban farming site identification in buildings}
\shortauthors{Ankit Palliwal et~al.}

\title [mode = title]{3D city models for urban farming site identification in buildings}                      

\author[1]{Ankit Palliwal} [orcid=0000-0002-2080-8352]
\address[1]{Department of Geography, National University of Singapore, Singapore}

\author[2]{Shuang Song} [orcid=0000-0001-8006-0537]
\author[2]{Hugh Tiang Wah Tan} [orcid=0000-0002-3056-4945]
\address[2]{Department of Biological Sciences, National University of Singapore, Singapore}

\author[3, 4]{Filip Biljecki}[orcid=0000-0002-6229-7749]
\cormark[1]
\ead{filip@nus.edu.sg}
\address[3]{Department of Architecture, National University of Singapore, Singapore}
\address[4]{Department of Real Estate, National University of Singapore, Singapore}
\cortext[cor1]{Corresponding author}

\begin{abstract}
Studies have suggested that there is farming potential in urban residential buildings. However, these studies are limited in scope, require field visits and time-consuming measurements. Furthermore, they have not suggested ways to identify suitable sites on a larger scale let alone means of surveying numerous micro-locations across the same building. Using a case study area focused on high-rise buildings in Singapore, this paper examines a novel application of three-dimensional (3D) city models to identify suitable farming micro-locations (level and orientation) in residential buildings. We specifically investigate whether the vertical spaces of these buildings comprising outdoor corridors, fa\c{c}ades and windows receive sufficient photosynthetically active radiation (PAR) for growing food crops and do so at a high resolution. We also analyze the spatio-temporal characteristics of PAR, and the impact of shadows and different weather conditions on PAR in the building. Environmental simulations on the 3D model of the study area indicated that the cumulative daily PAR or Daily Light Integral (DLI) at a location in the building was dependent on its orientation and shape, sun's diurnal and annual motion, weather conditions, and shadowing effects of the building's own fa\c{c}ades and surrounding buildings. The DLI in the study area generally increased with building's levels and, depending on the particular micro-location, was found suitable for growing moderately light-demanding crops such as lettuce and sweet pepper. These variations in DLI at different locations of the same building affirmed the need for such simulations. The simulations were validated with field measurements of PAR, and correlation coefficients between them exceeded 0.5 in most cases thus, making a case that 3D city models offer a promising practical solution to identifying suitable farming locations in residential buildings, and have the potential for urban-scale applications.
\end{abstract}




\maketitle

\begin{textblock*}{\textwidth}(0cm,-14.3cm)
\begin{center}
\begin{footnotesize}
\begin{boxedminipage}{1\textwidth}
This is the accepted manuscript of an article published by Elsevier in the journal \emph{Computers, Environment and Urban Systems}. The published journal article is available at: \url{https://doi.org/10.1016/j.compenvurbsys.2020.101584}. Cite as:
Palliwal A, Song S, Tan HTW, Biljecki F (2021): 3D city models for urban farming site identification in buildings. \textit{Computers, Environment and Urban Systems}, 86: 101584.
\end{boxedminipage}
\end{footnotesize}
\end{center}
\end{textblock*}

\begin{textblock*}{\textwidth}(-0.5cm,13cm)
{\tiny{\copyright{ }2021, Elsevier. Licensed under the Creative Commons Attribution-NonCommercial-NoDerivatives 4.0 International (\url{http://creativecommons.org/licenses/by-nc-nd/4.0/})}}
\end{textblock*}

\section{Introduction}\label{sec:intro}
\begin{figure*}[pos = htbp]

\begin{subfigure}{0.5\textwidth}
	\centering
	\includegraphics[width=0.9\linewidth, height=5cm]{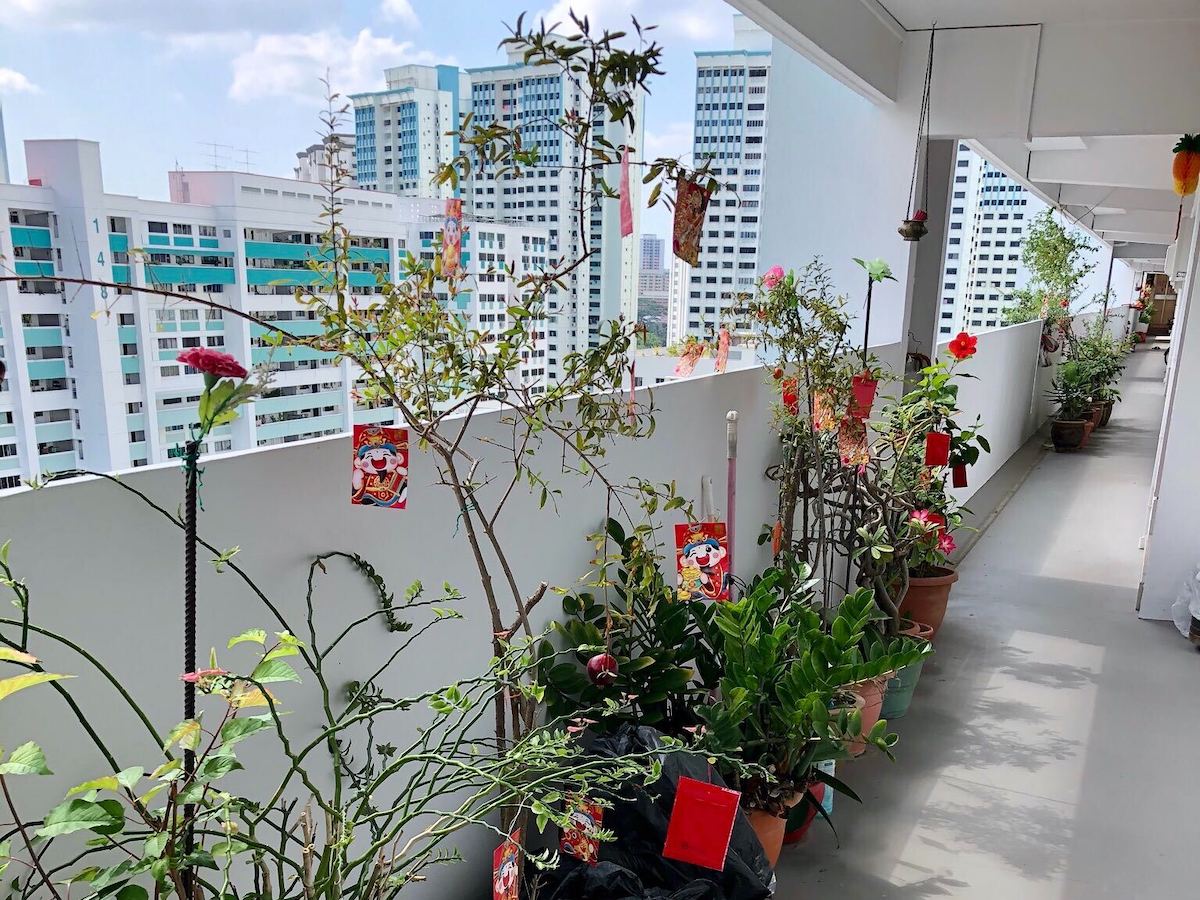}
	\caption{Corridors}
	\label{fig:corridors}
\end{subfigure}%
\begin{subfigure}{0.5\textwidth}
	\centering	
	\includegraphics[width=0.5\linewidth, height=5cm]{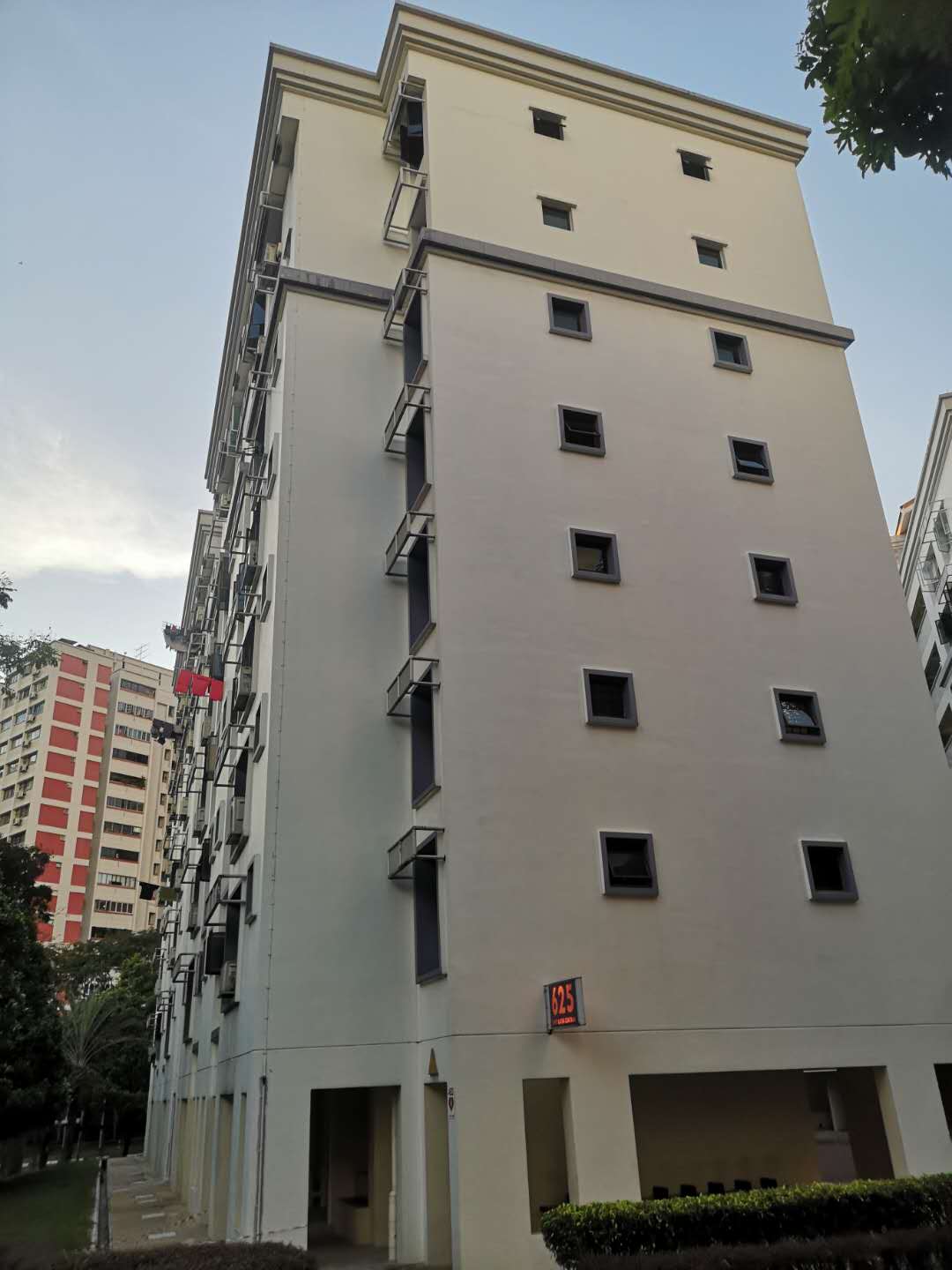}
	\caption{Fa\c{c}ades}
	\label{fig:facades}
\end{subfigure}

\begin{subfigure}{0.5\textwidth}
	\centering	
	\includegraphics[width=0.9\linewidth, height=5cm]{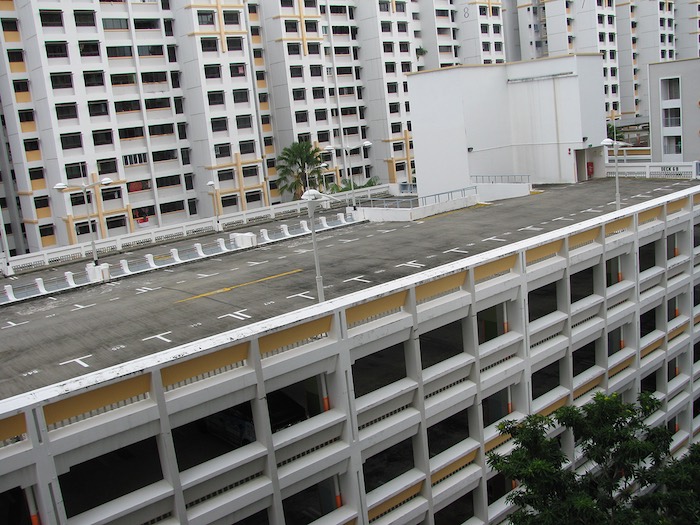}
	\caption{Rooftops}
	\label{fig:rooftops}
\end{subfigure}%
\begin{subfigure}{0.5\textwidth}
	\centering	
	\includegraphics[width=0.5\linewidth, height=5cm]{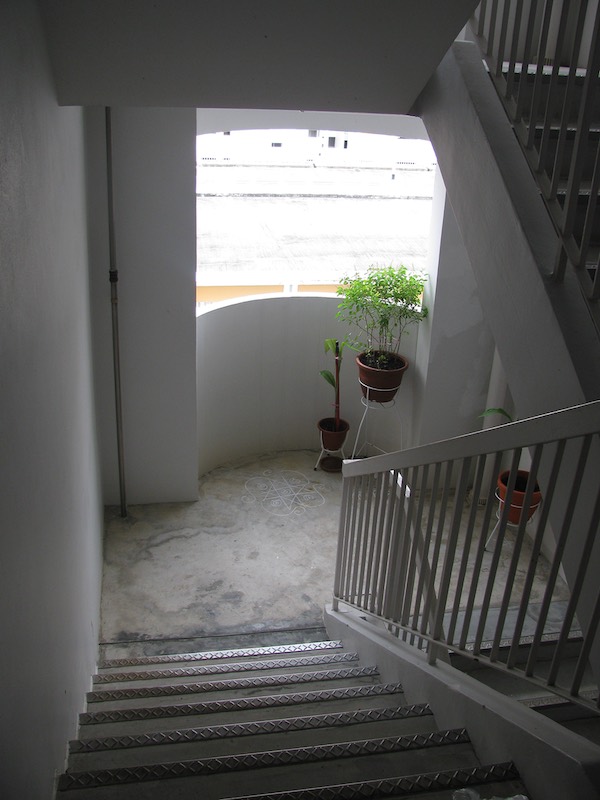}
	\caption{Staircases}
	\label{fig:staircases}
\end{subfigure}

\begin{subfigure}{0.5\textwidth}

	\includegraphics[width=0.9\linewidth, height=5cm]{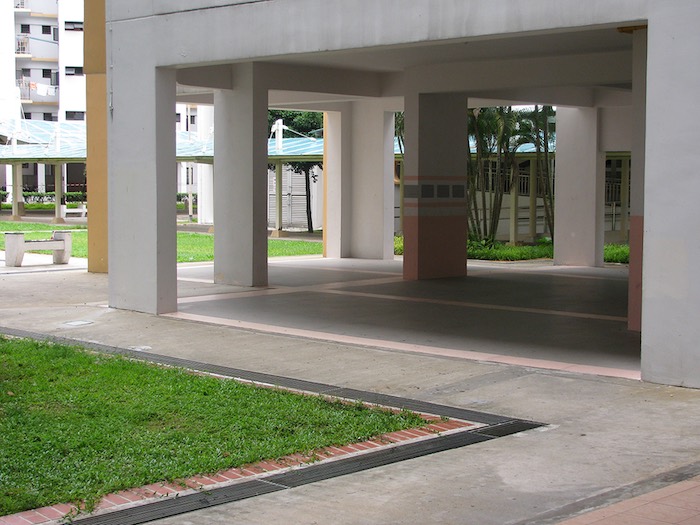}
	\caption{Void decks}
	\label{fig:voiddecks}
\end{subfigure}%
\begin{subfigure}{0.5\textwidth}
	\centering	
	\includegraphics[width=0.5\linewidth, height=5cm]{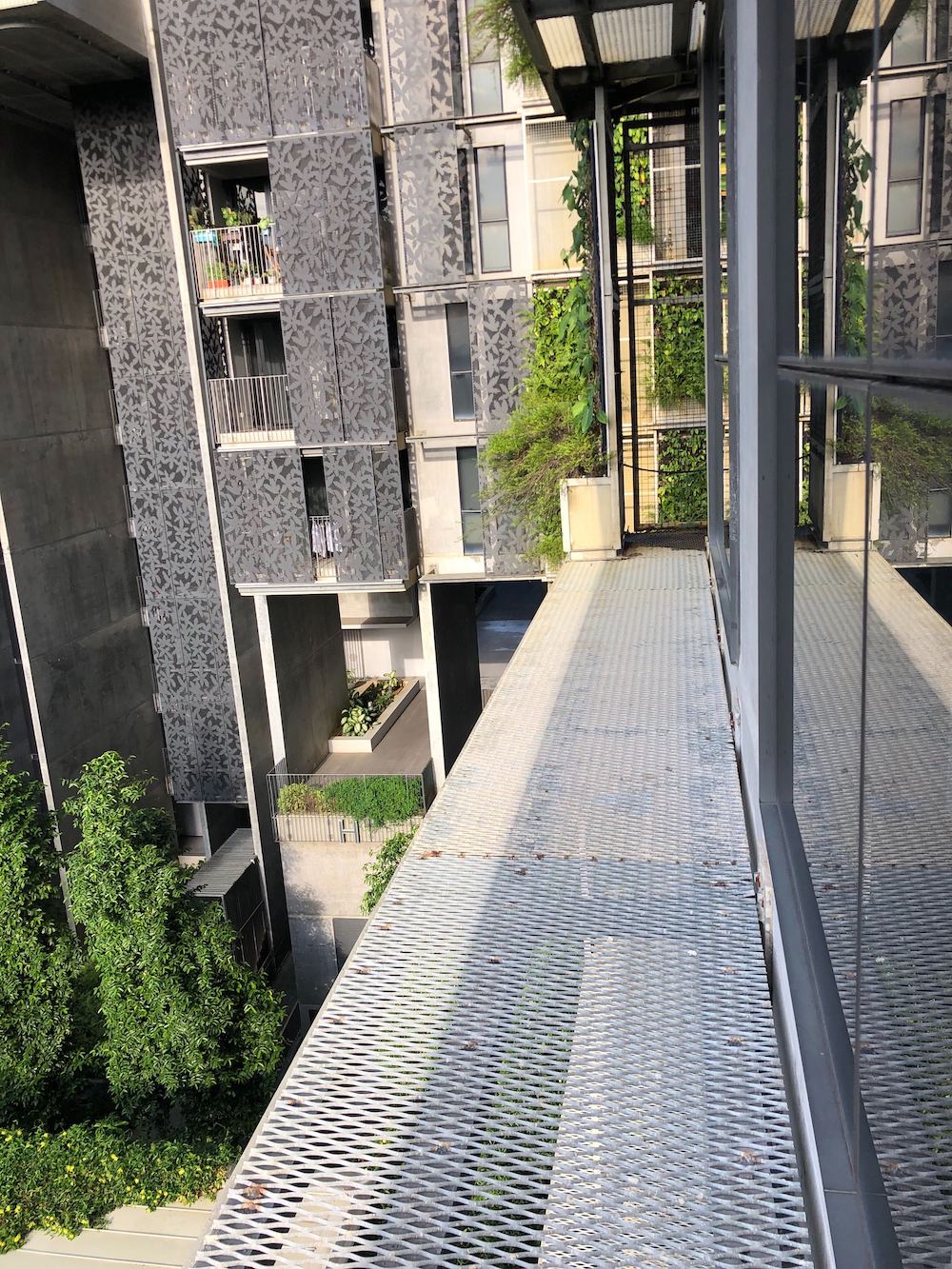} 
	\caption{Window ledges}
	\label{fig:windowledges}
\end{subfigure}

\caption{Open spaces in urban and high-rise residential buildings that may have farming potential. In this paper, we mainly focused on outdoor vertical spaces (Figures \ref{fig:corridors}, \ref{fig:facades}, and \ref{fig:windowledges}) since they are the sites that are most suitable for use in our study area (e.g.\ access to rooftops is usually restricted or reserved for other purposes), but our methodology is sufficiently generic that it can be applied to other parts of buildings and in different geographic locations.}
\label{fig:openspaces}
\end{figure*}

Over the years, farming in and around urban buildings, particularly residential buildings, has gained popularity in high-density and high-rise environments~\citep{lim2010,  khanetal2018, kimetal2018, songetal2018, kosoricetal2019}. This is primarily because with limited land available for agriculture, these buildings offer under-utilized horizontal and vertical spaces that may have farming potential (Figure~\ref{fig:openspaces}). In addition, improvement of emotional, mental and physical well-being of the occupants \citep{tanismail2015}, mitigation of the urban heat island effect~\citep{diehl2020}, creation of job opportunities~\citep{tablada2016}, and reduction in carbon emissions associated with transportation of food~\citep{lim2010} are counted among the other benefits of farming in these buildings. In Singapore, urban farming carries special significance as it has been adopted as one of the `Grow Local' strategies to achieve the `30 by 30' vision of the Singapore Food Agency (SFA)~\citep{zulkifli2019}. This vision aims to locally produce 30\% of Singapore's nutritional needs by 2030.

Situated at 1$^{\circ}$ North of equator, Singapore is an arable land-scarce and densely populated island city-state which accommodates a population of about 5.7 million~\citep{sdos2019b} over its land area of 722.5km$^{2}$~\citep{sdos2019a}. Having only 1\% of land set aside for agriculture~\citep{diehl2020}, Singapore meets 90\% of its food requirements through imports~\citep{kosoricetal2019} leaving it vulnerable to external food price fluctuations and disruptions in the food supply chain. To reduce this heavy reliance on food imports and inline with SFA's vision, Singapore's numerous high-rise residential buildings, which accommodate almost its entire population, emerge as promising sites for urban farming. These buildings not only offer under-utilized spaces but also provide an opportunity to its occupants, who are considered a key stakeholder to drive urban agriculture in Singapore, the ease of engaging in farming while at home \citep{kosoricetal2019}.
The vast majority of these buildings are public housing buildings, whose number surpassed 10 thousands and accommodate more than 80\% of the nation~\citep{hdb2020,kosoricetal2019}.
They have standardised designs and due to their public nature offer prospects for government-supported initiatives.

Besides the optimal crop growth conditions such as the level of carbon dioxide, nutrients, humidity, temperature, water, among others -- which are fairly constant at the building scale -- successful identification of soil-based farming micro-locations (particular site at a specific level) in a building requires an assessment of its exposure to sunlight. Past studies have measured sunlight availability either in terms of its duration or its level. The former approach mainly focuses on the minimum duration of direct sunlight required by the crops for their growth. Based on the direct sunlight duration, suitable crops are identified for growing at particular sites \citep{wongandlau2013, santosetal2016, kimetal2018}. While this approach may be useful for estimating urban green areas at large scale using remote sensing data (e.g. \cite{santosetal2016}), it may not be suitable for identifying potential urban farming sites as different crops are sensitive to different levels of sunlight referred to as photosynthetically active radiation (PAR) \citep{tanismail2015}. PAR forms the basis for the latter approach and is defined as the portion of solar spectrum, in the 400 to 700 nm wavelength range, that is utilized by plants for photosynthesis. Its amount is a key factor to understand whether there is a potential for farming and what kind of crops can be grown at a specific site because different crops require different PAR conditions for its optimal growth~\citep{songetal2018}. Conventionally, PAR assessment in different urban forms is accomplished through field surveys that involves placing PAR sensors at selected locations such as a few spots in a building~\citep{tanismail2014, tanismail2015, songetal2018}. The findings of these surveys suggest that, in a high density urban environment, different urban forms are exposed to different levels of PAR due to their shape, orientation, self-shadowing effects, and shadowing effects of surrounding objects~\citep{tanismail2014}. Furthermore, PAR at a given micro-location also varies due to changes in the sun's position in the sky and different weather conditions~\citep{songetal2018}, thus, highlighting the need to understand the spatio-temporal characteristics of PAR at the \textit{building's micro-locations}. While these studies confirm that there is farming potential in residential buildings, they are: (i) constrained -- only a limited number of locations can be covered with sensors, and as this paper will show, there might be a large variation of PAR even at the same side of a building between different levels; and (ii) arduous -- as the surveys require field visits and the sensors have to be installed for an extended time period such as weeks rather than enabling instantaneous measurements.
Furthermore, they have not suggested ways on how to possibly go about estimating the potential at the urban scale.

This paper investigates whether three-dimensional (3D) city models can be used to assess the suitability of particular micro-locations in high-rise buildings for urban farming, leading to bypassing building visits and measurements while taking into account the peculiarities associated with sunlight availability in built environments, and calculating the potential of unlocking under-utilized spaces in residential buildings for urban farming.
Our work capitalizes on the rich body of knowledge on using 3D city models for understanding the benefit of installing solar panels in buildings (Section~\ref{sec:litreview}) and adapts the work to enable simulations suited for gathering the potential of urban farming (Section~\ref{sec:methods}).
The study has run environmental simulations to analyze the spatio-temporal characteristics of PAR, assess adequacy of PAR received for growing crops, and understand the influence of different weather conditions, self-shadowing, and shadowing effects of nearby urban forms.
Unlike the vast majority of papers dealing with simulations in 3D GIS, we conduct field measurements to verify the veracity of the simulations and conclude that 3D city models are a viable instrument for calculating the potential of spaces in buildings for urban farming (Section~\ref{sec:results}).
To the best of our knowledge, 3D city models have not been used for this purpose before.
Our results are important because conducting field visits and undertaking PAR measurements to identify all locations in these buildings that receive adequate sunlight for growing crops can prove to be a difficult task, and the work can lead to estimations of the urban farming potential at the precinct or at the urban scale, enabling future studies considering thousands of buildings at once, similar to other applications in (3D) GIS. 
In this complete process of employing 3D city models from acquisition all the way to analysis and extraction of insights, this paper also presents an alternative method of estimating building heights in the absence of conventional data by measuring staircases, which has not been documented in the existing academic literature.

\section{Literature review}\label{sec:litreview}
Geospatial technologies have been applied for long across diverse agricultural and allied activities such as precision farming~\citep{wilson2005}, assessment of land suitability for agriculture~\citep{bandyo2009}, suitability analysis for beekeeping sites~\citep{estoque2011}, and more recently quantification of potential green cover on rooftops~\citep{santosetal2016}. However, hitherto 3D geoinformation has not been used to identify urban farming sites despite their wide usage to assess the availability of solar energy in built environments for installing photovoltaic panels on buildings~\citep{redweiketal2013, catitaetal2014, freitas2015, rubioetal2016, sarettaetal2020}.
These simulations, which are primarily focused on rooftops of buildings, are used to determine whether a particular part of a building/rooftop receives sufficient solar exposure to warrant the installation of a solar panel.
The amount of solar exposure is primarily influenced by the geographic location, orientation, and nearby objects that cause (self-)shadowing.
The fact that 3D city models provide sufficient information for such simulations catalyzed the development of this long-standing research line.

Since urban farming much depends on the available level of light, which directly dictates the suitability of particular types of crops and influences the agricultural yield of crops (as much as it drives the energy yield of solar panels), our work takes advantage of the developments in the energy department, and seeks into leveraging them for a different purpose essentially establishing a new research line marrying urban agriculture and 3D GIS.

Given that an integral component of our work is generating a 3D model of the study area, it is worthwhile to provide a short literature review of the process.
Most 3D city models are generated by extrusion, combining building footprints and data on building heights, usually obtained from lidar point clouds~\citep{dukai2019}.
This process results in block building models (or LOD1 as per CityGML/CityJSON~\citep{Groger:2012hg,Ledoux:2019tx}), which despite their coarse nature have proven useful in scores of simulations such as predictions of the impact of noise in the built enviroment~\citep{stoter2020}.

However, point clouds, a reliable but expensive source of building heights, are often unavailable, as it is the case for our study area.
To counter this gap, alternative methods of deriving building heights in absence of direct elevation measurements have emerged.
\cite{biljeckietal2017} review several of them concluding that the most common unorthodox approach is using the number of levels of a building as a proxy for its height, which continues to be engaged in many studies, such as for energy simulations~\citep{Cheng:2020dm}.
Another method, recently published in this journal, demonstrates that heights can be estimated from a single photograph captured through smartphones~\citep{bshoutyetal2020}.
While for our study area we have at disposal a 3D city model generated using open data on building levels, which is reasonably accurate, we take advantage of the fact that all buildings there are publicly accessible (being public housing buildings) and count the number of stairsteps across their vertical extent.
As trivial as this approach appears, we believe that it is powerful, presenting another contribution of ours, which may warrant attention for future investigations, especially in the context of crowdsourcing building heights.

\section{Materials and methods}\label{sec:methods}
\subsection{Study area}

\begin{figure}[pos = htbp]
	\centering
	\includegraphics[width=0.5\textwidth]{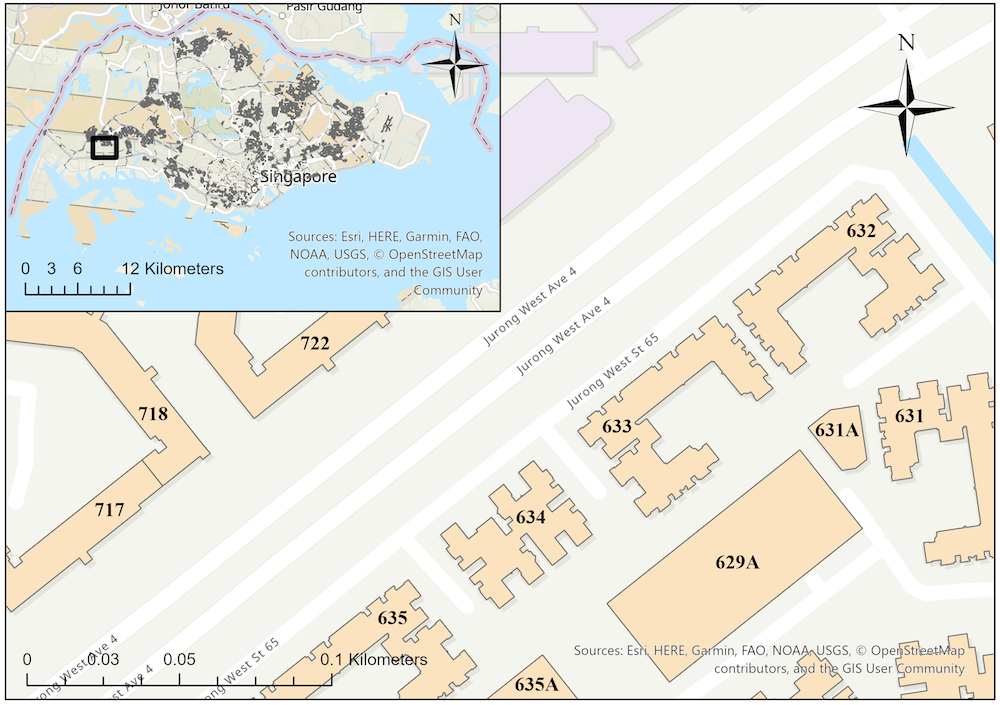}
	\caption{\label{fig:studyArea}Study area: Block 633, Jurong West Street 65, Singapore.}
\end{figure}

\begin{figure*}[pos = htbp]
\begin{subfigure}{0.5\textwidth}
	\centering
	\includegraphics[width=0.9\linewidth, height=5cm]{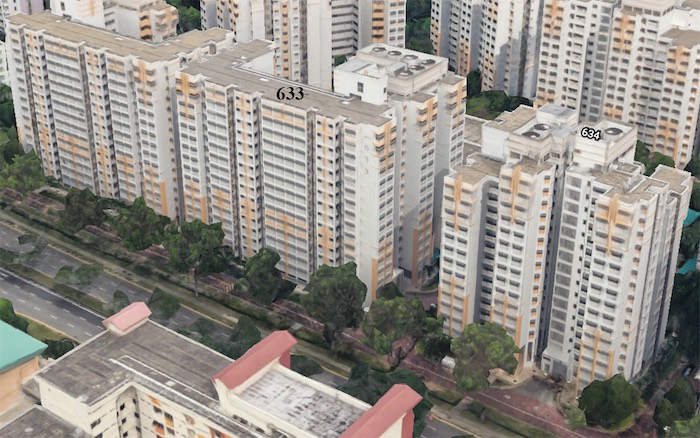} 
\end{subfigure}%
\begin{subfigure}{0.5\textwidth}
	\centering	
	\includegraphics[width=0.9\linewidth, height=5cm]{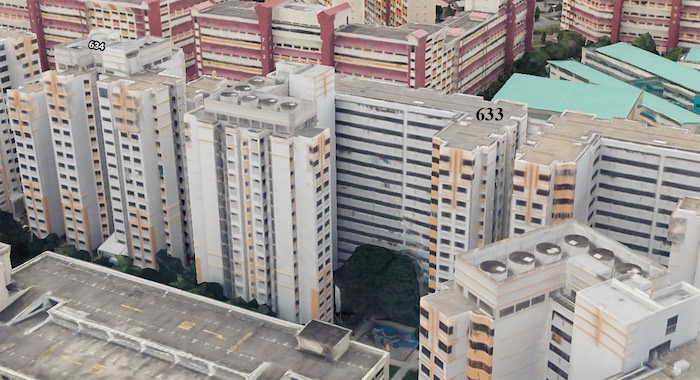}
\end{subfigure}
\caption{Buildings surrounding the public housing block in focus~\citep{gmaps2020a, gmaps2020b}.}
\label{fig:urbanForms}
\end{figure*}

The study area consisted of the buildings situated on Jurong West Street 65 in Singapore (Figure \ref{fig:studyArea}). Jurong West is a residential town in the West region of Singapore. For the analyses, we focused on the fa\c{c}ades having corridors and window ledges of public housing building `Block 633'. According to~\cite{kosoricetal2019}, these micro-locations are the most preferred for farming among the occupants, presenting an appropriate focus. The construction of `Block 633' was completed in 2000. This residential building has 138 dwelling units spread over 16 levels~\citep{hdb2020}, and it shares the same design as many other public housing buildings across Singapore, rendering our study generic and not constrained to a particular building. Three of its sides have other residential buildings and a multi-storey car park adjoining them. The fourth side has a school and another residential building across the street (Figure \ref{fig:urbanForms}).
The dense built form inevitably results in shadowing, compounding the uncertainty of the amount of solar exposure required for urban farming and cultivating particular crops.

\subsection{Dataset}\label{sc:dataset}

\begin{figure}[pos = htbp]
	\centering
	\includegraphics[width=0.5\textwidth]{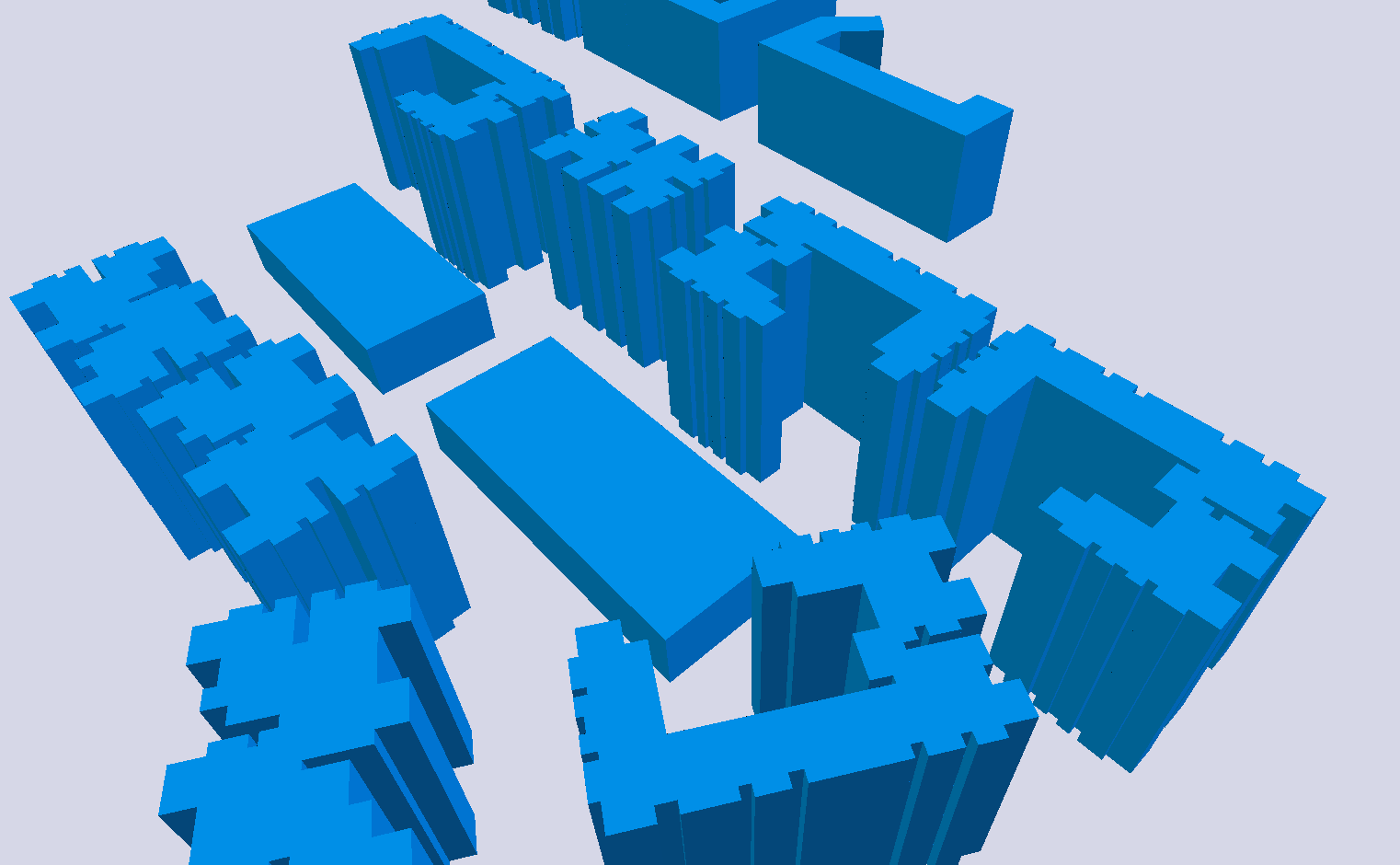}
	\caption{\label{fig:3Dmodel}3D model of the study area, generated from open data and using open-source software.}
\end{figure}

A 3D city model of the study area is available as open data~\citep{biljecki2020}.
This dataset has been generated combining building footprints available in OpenStreetMap with the number of levels released by the public housing agency.
In order to double down on the accuracy of the data, we have investigated whether there are alternative approaches to estimate the building heights, which would be somewhat more accurate than using the number of levels as a proxy.
To estimate building heights, the number of stairsteps from ground level/floor to the top level were counted, for each building around the block in focus. The floor-to-floor height was obtained by multiplying the stairstep count between two consecutive levels with the measured stairstep height. The derived floor-to-floor height was in consonance with the typical level one and floor-to-floor heights of 3.6m and 2.8m respectively in these buildings~\citep{hdb2014}. Finally, the building height was estimated by summing the floor-to-floor heights across all levels in the building~\citep{hdb2020}.
The generated 3D city model of the study area is shown in Figure~\ref{fig:3Dmodel}.

\subsection{Methodology, simulations, and tools}

The methodology consists of conducting solar exposure estimations adapted for urban agriculture and vertical spaces, and carrying out conventional measurements to verify the results.

For the simulations, Blender v2.79a and VI-Suite v0.4 have been used. VI-Suite is a free open source add-on package for Blender, a popular 3D computer graphics software. It consists of building environment performance simulation modules that allow 3D geospatial data analysis and visualization~\citep{southall2017}. It has the ability to (1) process large 3D geospatial datasets through a user-friendly interface, (2) integrate customized Python scripts, and (3) export the simulation results in comma separated values format for subsequent analysis; which made it suitable for this research and which we have done transporting the results to R v3.6.3.
It is important to note that these tools are free and open-source, so together with the open data (Section~\ref{sc:dataset}), it means that our work relies entirely on open sources, facilitating its reproducibility.

This study has used VI-Suite modules for sun path, shadow map, and lighting analysis. While sun path analysis displays the sun's position and its trajectory relative to the 3D model at any date, time and location; shadow mapping, on the other hand, calculates the percentage of time of the simulation period a location was exposed to direct sunlight on a sunny day. For these analyses, VI-Suite uses some of the in-built functionalities of Blender \citep{southall2017}. With lighting analysis, it is possible to calculate the irradiance values at discrete moments in time (referred to as basic lighting in VI-Suite) as well as the cumulative solar radiation received at a location over the simulation period (also known as Climate Based Daylight Modelling (CBDM)). For lighting analysis, VI-Suite uses Radiance lighting simulation suite in the background. Radiance is based on a backward ray-tracing daylight simulation method and is considered among the best freeware available for daylighting analysis \citep{ward1994, freitas2015}.

\begin{figure*}[pos = htbp]

\begin{subfigure}{0.5\textwidth}
	\centering
	\includegraphics[width=0.9\linewidth, height=5cm]{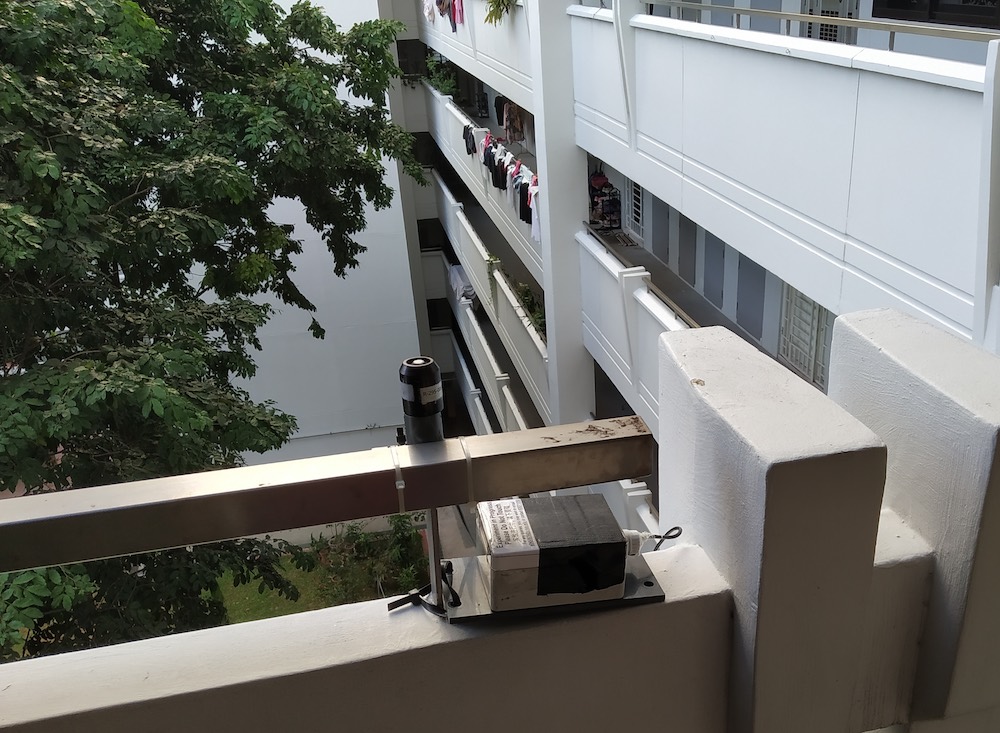} 
	\caption{PAR sensor placement along the corridors.}
	\label{fig:placementCorridor}
\end{subfigure}%
\begin{subfigure}{0.5\textwidth}
	\centering	
	\includegraphics[width=0.9\linewidth, height=5cm]{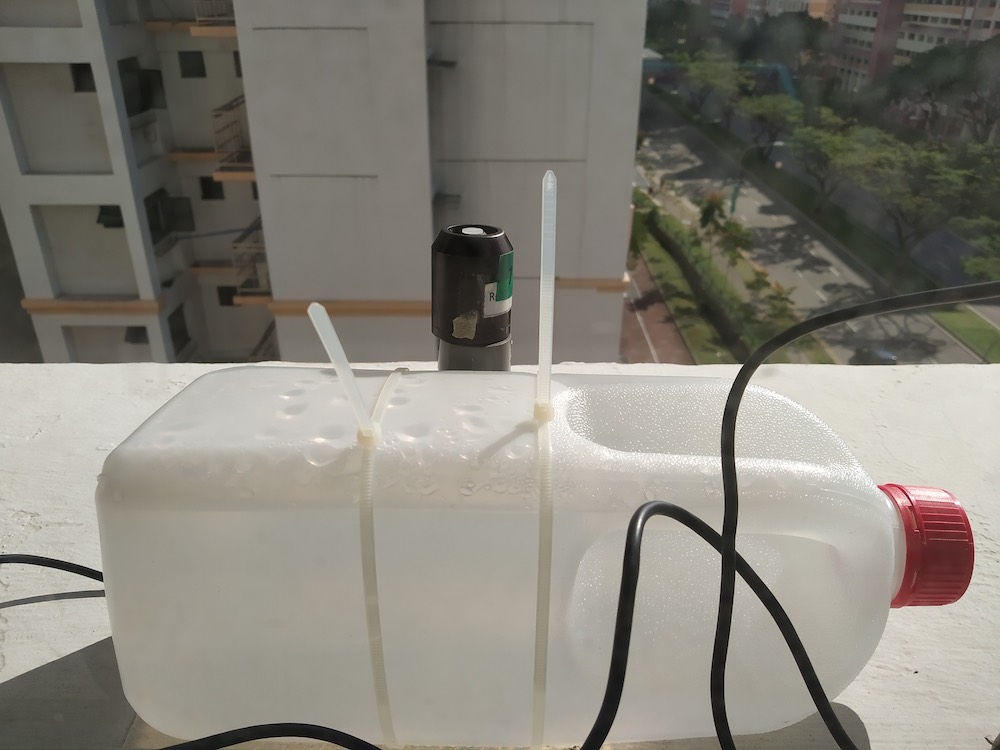}
	\caption{PAR sensor placement on the window ledge.}
	\label{fig:placementWindow}
\end{subfigure}

\begin{subfigure}{0.9\textwidth}
	\centering	
	\includegraphics[width=0.5\linewidth, height=5cm]{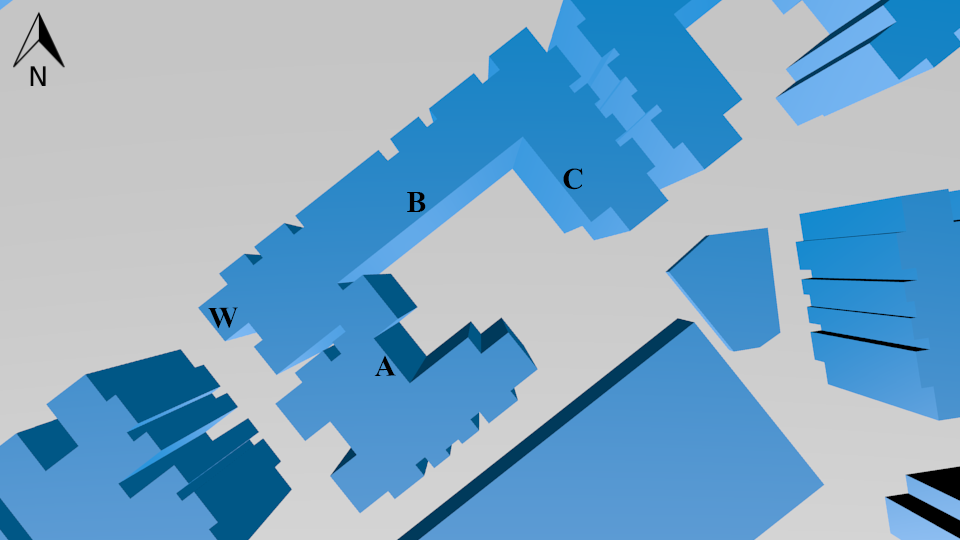}
	\caption{Fa\c{c}ade description in the study area on the 3D model.}
	\label{fig:facadeDesc}
\end{subfigure}
\begin{subfigure}{0.9\textwidth}
	\centering	
	\includegraphics[width=0.9\linewidth]{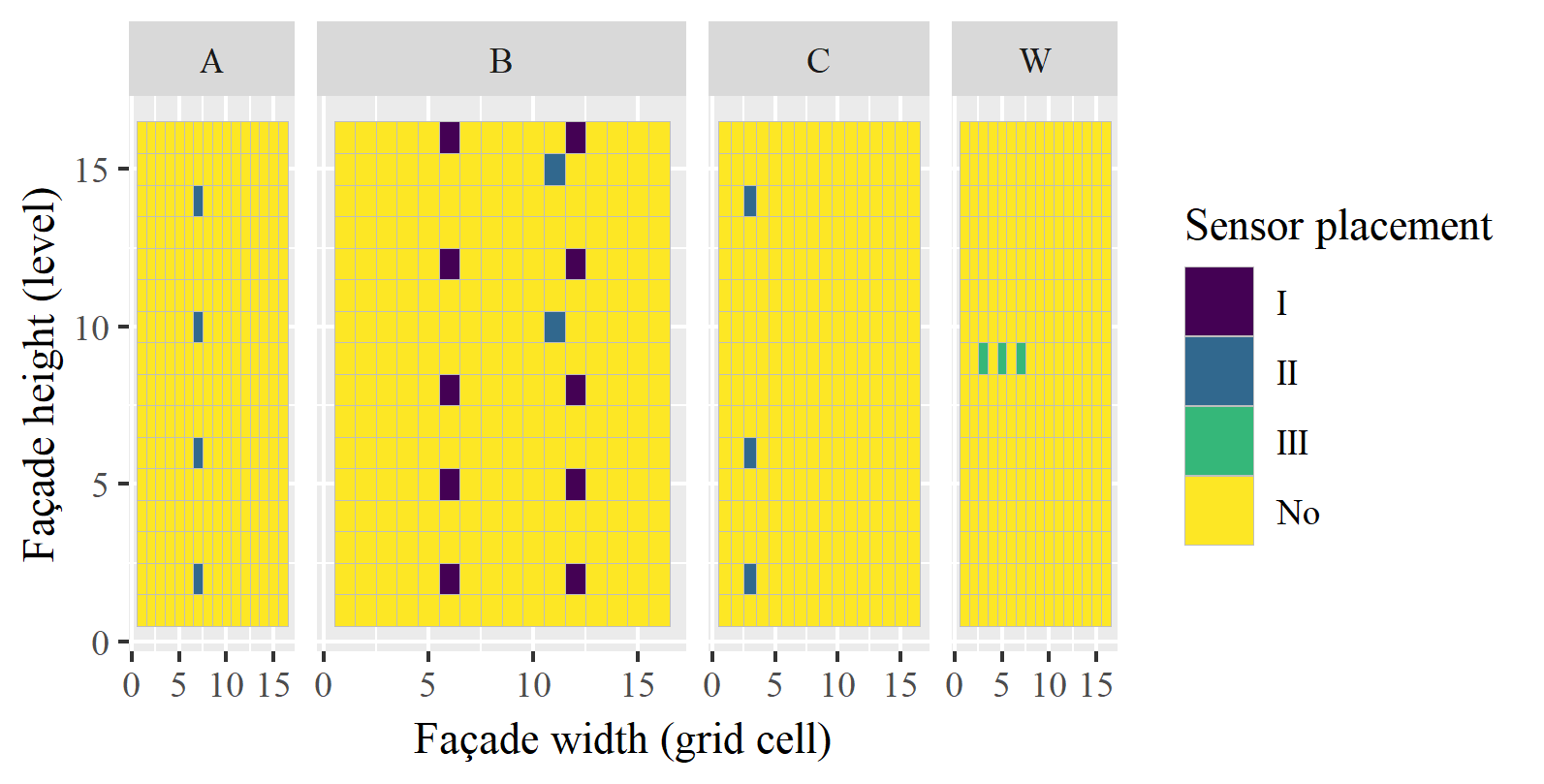}
	\caption{Sensor locations on the gridded fa\c{c}ades of the 3D model during survey periods.}
	\label{fig:sensor}
\end{subfigure}

\caption{Fa\c{c}ade description and sensor placement in the study area during survey periods.}
\label{fig:descPlacement}
\end{figure*}

This research mainly focused on environmental simulations during three periods: (I) 02 Mar 2020 6am -- 11 Mar 2020 6am, (II) 15 Mar 2020 6am -- 25 Mar 2020 6am, and (III) 27 Mar 2020 6am -- 06 Apr 2020 6am, accompanied by PAR surveys conducted in the study area. Instantaneous PAR was measured at several locations along the corridors of the residential building (I, II) and one of the window ledges of a dwelling unit (III). PAR was measured using Onset PAR Smart Sensor (S-LIA-M003) with a sampling interval of one second and mean values logged at five-minute intervals in HOBO Micro Station (H21-USB and H21-002) data logger. Prior to conducting survey, these sensors were calibrated against an Odyssey PAR logger according to the manufacturer's manual\footnote{The respective calibration equations for the sensors along with coefficients of determination (R$^{2}$) can be found in Table \ref{tbl1} in Appendix \ref{sec:appendix}.}. These sensors were placed on various levels of the building along the corridor railings of fa\c{c}ades A, B, and C and on one of the window ledges of fa\c{c}ade W (Figures \ref{fig:placementCorridor}, \ref{fig:placementWindow}, and \ref{fig:facadeDesc}). Sky conditions/daily weather forecasts were also monitored during these periods \citep{weatherchannel2020, weather2020}.

The precise location of each sensor placed in the building was determined using a measuring tape. To match these locations in the 3D model, each fa\c{c}ade of the 3D model in Figure \ref{fig:facadeDesc} was converted into a 16$\times$16 grid (Figure \ref{fig:sensor}\footnote{The numbers 1-16 on the vertical and horizontal axes in this figure represent the $i^{th}$ level and grid cell respectively.}). Matching the positions of sensors to the corresponding grid cells in the 3D model is essential to enable comparisons.
Height and width of the fa\c{c}ades in the 3D model were determined from the local coordinates of their vertices. Based on proportionality and determined height and width of the fa\c{c}ades, each sensor location in the building was mapped to the corresponding grid cell in the 3D model (Figure \ref{fig:sensor}). To illustrate, during survey period I, sensors were placed in grid cells 6 and 12 at levels 2, 5, 8, 12, and 16 of fa\c{c}ade B.

\begin{figure}[pos = htbp]
	\centering
	\includegraphics[width=0.5\textwidth]{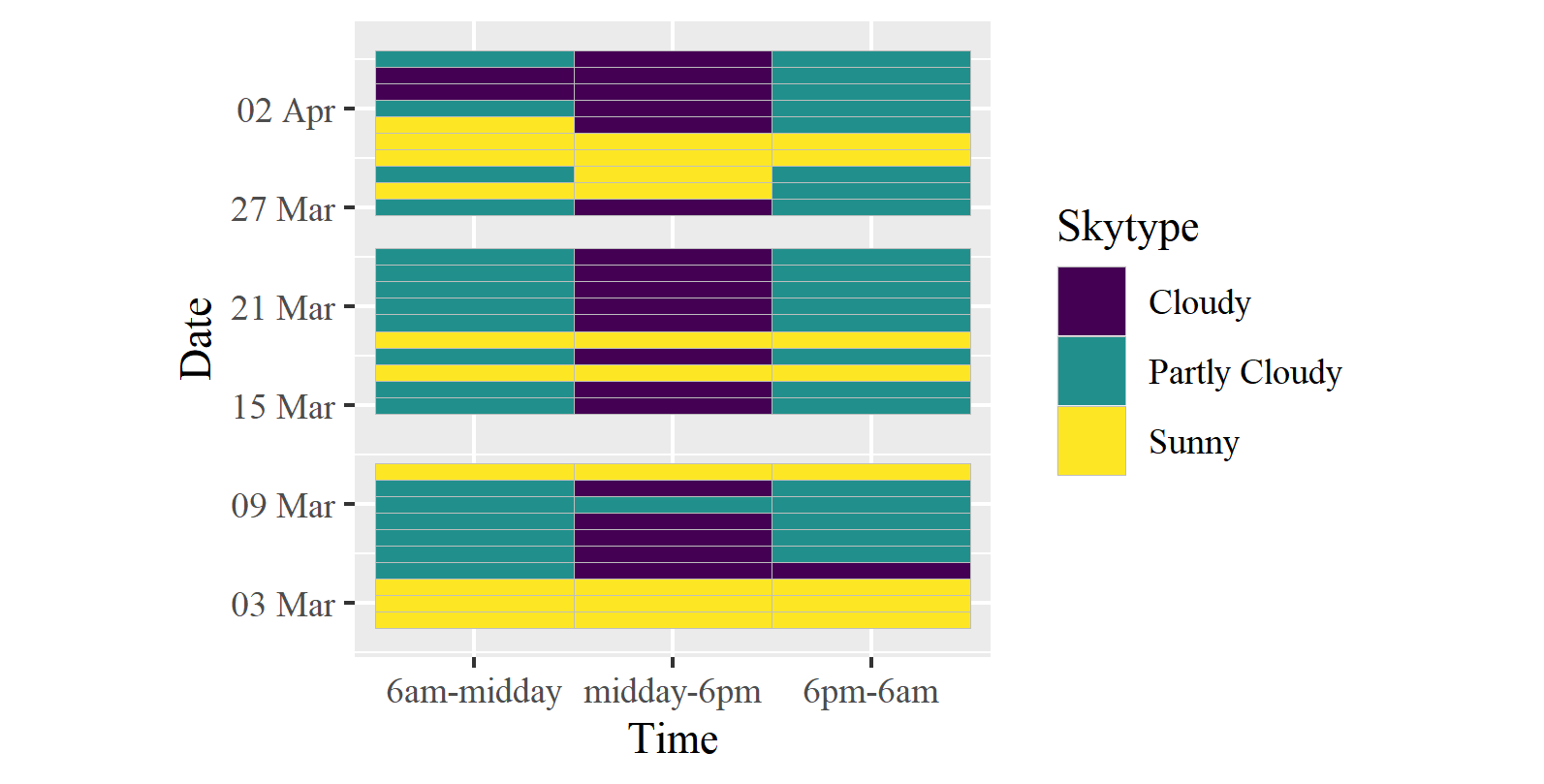}
	\caption{\label{fig:skytype}Skytype based on 24-hr weather forecast during survey periods.}
\end{figure}

Running simulations in VI-Suite requires one or more of the following as inputs: (1) latitude/longitude information of the study area, (2) weather data of the study area in EnergyPlus Weather (EPW) format, (3) sky/weather condition (called skytype in VI-Suite) namely, sunny, partly cloudy, and cloudy during simulation period, and (4) start date/hour and end date/hour of the simulation period.
This study used the weather data of Singapore available on EnergyPlus website \citep{ashrae2001} which has also been used in \cite{tanismail2014, tanismail2015}. Skytype during survey periods (Figure \ref{fig:skytype}) were based on National Environment Agency's 24-hr weather forecast for the West region \citep{weather2020}. This forecast is typically available for morning (6am--midday), afternoon (midday--6pm), and night (6pm--6am). In these forecasts, `Fair' and `Fair \& Warm' sky conditions were classified as `Sunny' and `Showers' and `Thundery Showers' sky conditions were classified as `Cloudy'. 

Start date/hour and end date/hour of the simulations were decided based on the objectives of this study. Sun-paths were generated for the first (02 Mar) and the last day (05 Apr) of the PAR survey. Rendered images from solar illumination at 10am and 1pm were also produced for these days to show the movements of shadows casted by buildings during and between these days. Shadow maps of fa\c{c}ades were generated for a sunny day (17 Mar) of the survey period. Maps were produced at discrete moments in time (10am, 1pm, and 4pm) and for different time periods (7am--1pm, 1pm--7pm, and 7am--7pm) to highlight the shadowing effects during the day. PAR simulations were carried out, using basic lighting analysis, at 10am, 1pm, and 4pm on a sunny (17 Mar), partly cloudy (09 Mar), and cloudy (03 Apr) day to analyze the spatio-temporal distribution of PAR on the fa\c{c}ades under different weather conditions. PAR simulations without the ground plane were also carried out for these time instances on the sunny day to demonstrate the effects of ground reflections. Simulated PAR (in $\mu$mol m$^{-2}$ s$^{-1}$ ($\Psi$)) were obtained by multiplying the solar irradiance (in W m$^{-2}$) from VI-suite by 2.02 \citetext{\citealp[p.36]{mavi2004}; \citealp[p.257]{foken2017}}.
Measurement units are one of the principal difference in comparison to studies focused on assessing the suitability of installing photovoltaic panels.
For assessing the level of sunlight availability, the results from such simulations cannot be used directly, but have to be converted to other units, which are not supported by simulation software.

While PAR refers to the instantaneous amount of solar radiation, according to \cite{songetal2018}, the adequacy of sunlight for crop growth is expressed in terms of DLI which is defined as the cumulative instantaneous PAR over 24-hour period. Thus, DLI (in mol m$^{-2}$ day$^{-1}$ ($\Phi$)) can be calculated for each sensor location in the study area using the equation: 

\begin{math}
	\text{DLI} = \sum_{i=1}^{288} (i^{th} \text{mean logged instantaneous PAR } ({\text{in } \Psi} ) \times 5 \times 60) \times 10^{-6}
\end{math}

where, $i$ depicts the $i^{th}$ five-minute interval in 24-hour period. For each location, average DLI can also be derived from these DLI for each survey period. The DLI equivalent in VI-Suite was obtained by using CBDM analysis wherein the hourly beam and diffuse solar radiation data of Singapore was taken in EPW format~\citep{southall2017}. For each grid cell, cumulative solar radiation (expressed in kWh m$^{-2}$) was obtained for the whole year as well as for the months of March, June, September, and December. Average values were obtained for each grid cell by dividing these cumulative values by the corresponding number of days in the month/year. Finally, simulated DLI (in $\Phi$) was obtained by multiplying the averaged values with 7.272 (1 kWh m$^{-2}$ day$^{-1}$ = 3.6 × 2.02 $\Phi$). These simulated DLI were rounded off to the nearest integer and then compared with the known DLI of different crops~\citep{faustnd, songetal2018} to identify the suitable crops for a given location. Lastly, hourly measured and simulated PAR from 7am to 7pm were obtained for each sensor location during the survey periods. Spearman's correlation coefficient ($\rho$) was determined between them due to their non-normal distributions. They were also compared based on mean absolute error (MAE) and root mean square error (RMSE). To enable reproducibility of the work, other parameter values used in the simulations of sun path, shadow map, and lighting analysis of VI-Suite have been included in this paper (see Table \ref{tbl2} in Appendix \ref{sec:appendix}).

\section{Results and discussion} \label{sec:results}

\subsection{Results}

The main result of the work is that 3D city models appear to be a promising tool for assessing the potential of urban farming in high-rise buildings and for identifying suitable sites.
Further, simulations using 3D city models can help to understand which crop is best suited for a particular site in a building.
This section elaborates on the results in details.

\subsubsection{Sun path analysis}
\begin{figure*}[pos = htbp]

\begin{subfigure}{0.9\textwidth}
	\centering
	\includegraphics[width=0.9\linewidth]{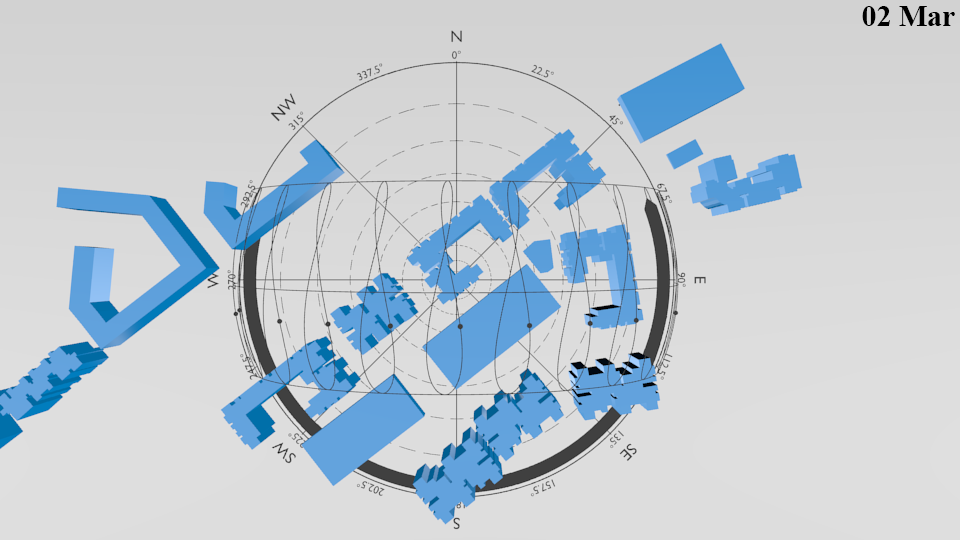} 
\end{subfigure}

\begin{subfigure}{0.9\textwidth}
	\centering	
	\includegraphics[width=0.9\linewidth]{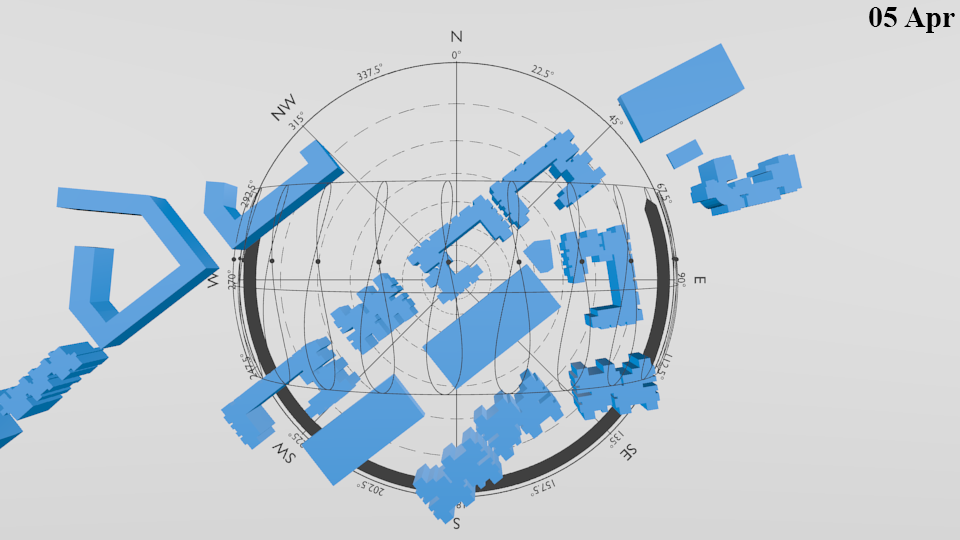}
\end{subfigure}

\caption{Hourly sun path diagrams on first and last day of the PAR survey.}
\label{fig:sunPath}
\end{figure*}

Figure \ref{fig:sunPath} shows the hourly sun path diagrams, which illustrates the sun's movement at different hours, on the first and last day of the PAR survey. The convoluted rings ($\infty$) depict the hours of the day from dawn to dusk with central ring representing noon. The points ($\bullet$) depict the sun's position at the hour represented by the ring.
In agreement with Figure \ref{fig:studyArea}, this figure also indicates that `Block 633' is oriented in the North West - South East direction.
As a result, the sunlight distribution is uneven on different fa\c{c}ades of the building owing to the sun's movement from East to West from dawn to dusk respectively. While fa\c{c}ades A and B that face North East and South East respectively received direct sunlight from morning till afternoon, fa\c{c}ades C and W that both face South West received direct sunlight during the afternoon and evening hours.

It can also be seen from the figure that the sun moved in the northern direction along the convoluted rings from 02 Mar to 05 Apr. In a typical year, the sun traverses from the southern extreme of this ring to the northern extreme during the first half (i.e. January--June) and in the reverse direction during the second half (i.e. July--December). This movement of the sun suggests a variation in PAR received at a given location. However, this variation during the year may not be significant in the context of this research due to the higher solar elevation given the equatorial position of Singapore \citep{tanismail2015, tablada2016}.

\begin{figure*}[pos = htbp]

\begin{subfigure}{0.5\textwidth}
	\centering
	\includegraphics[width=0.9\linewidth]{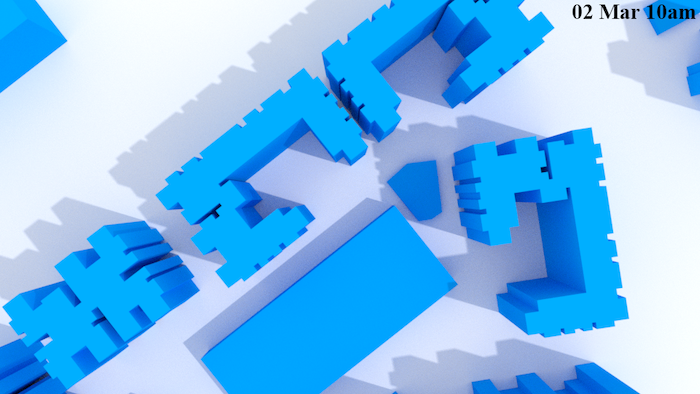} 
\end{subfigure}%
\begin{subfigure}{0.5\textwidth}
	\centering	
	\includegraphics[width=0.9\linewidth]{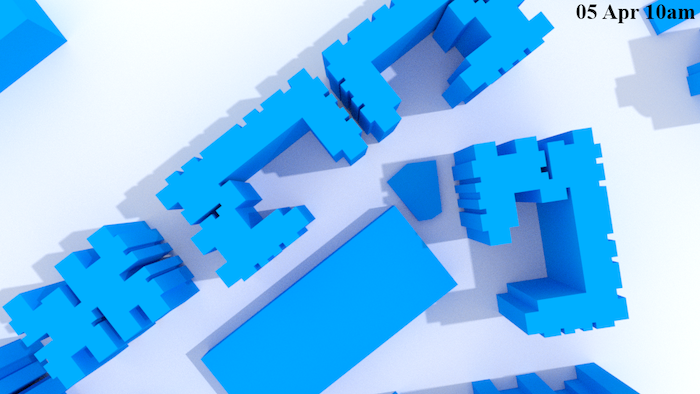}
\end{subfigure}

\begin{subfigure}{0.5\textwidth}
	\centering
	\includegraphics[width=0.9\linewidth]{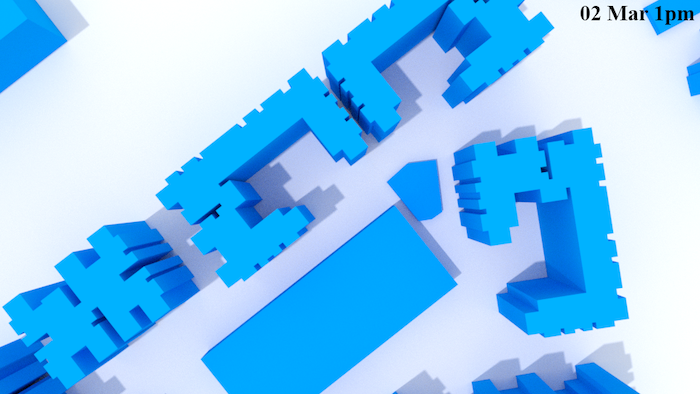} 
\end{subfigure}%
\begin{subfigure}{0.5\textwidth}
	\centering	
	\includegraphics[width=0.9\linewidth]{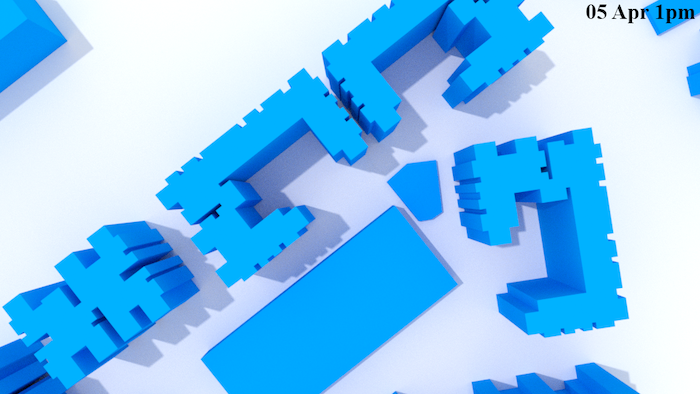}
\end{subfigure}

\caption{Solar illumination at different hours of the first and last day of survey.}
\label{fig:solarIllum}
\end{figure*}

Figure \ref{fig:solarIllum} shows the role played by shadowing effects on a day-to-day basis. It can be seen from this figure that PAR received on different fa\c{c}ades of the buildings was affected by the shadows casted by the building's own fa\c{c}ades and surrounding buildings. Further, the size of these shadows varied at different hours of the day owing to the sun's diurnal motion from East to West. For example, fa\c{c}ade B is partly shadowed by fa\c{c}ade C and shadow casted by a nearby building in the morning and as the day progresses, these shadowing effects recede. In addition, the size and the orientation of these shadows are affected by the annual motion of the sun which can be observed from the buildings' shadows in the figure on the first and last day of the PAR survey. For example, the size of the shadow casted by fa\c{c}ade C on fa\c{c}ade B at 10am is relatively larger on 05 Apr as compared to 02 Mar. Moreover, the shadows casted also depend on the shape of the building. For instance, shadow casted by `Block 633' on itself and nearby buildings differs from the shadows casted by nearby buildings of different shapes in the study area.    

\subsubsection{Shadow map analysis}

\begin{figure*}[pos = htbp]

\begin{subfigure}{0.5\textwidth}
	\centering
	\includegraphics[width=0.9\linewidth]{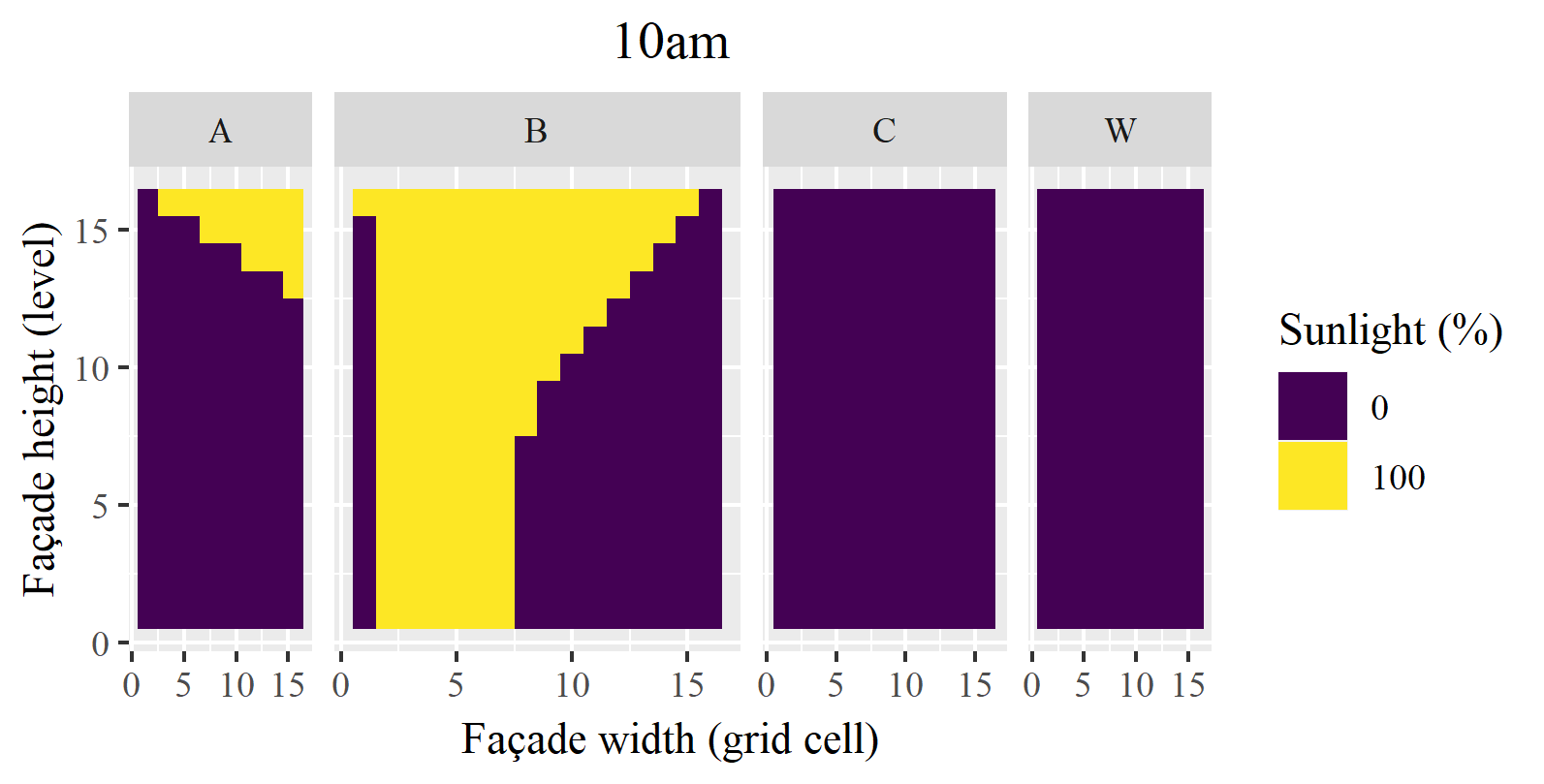} 
\end{subfigure}%
\begin{subfigure}{0.5\textwidth}
	\centering	
	\includegraphics[width=0.9\linewidth]{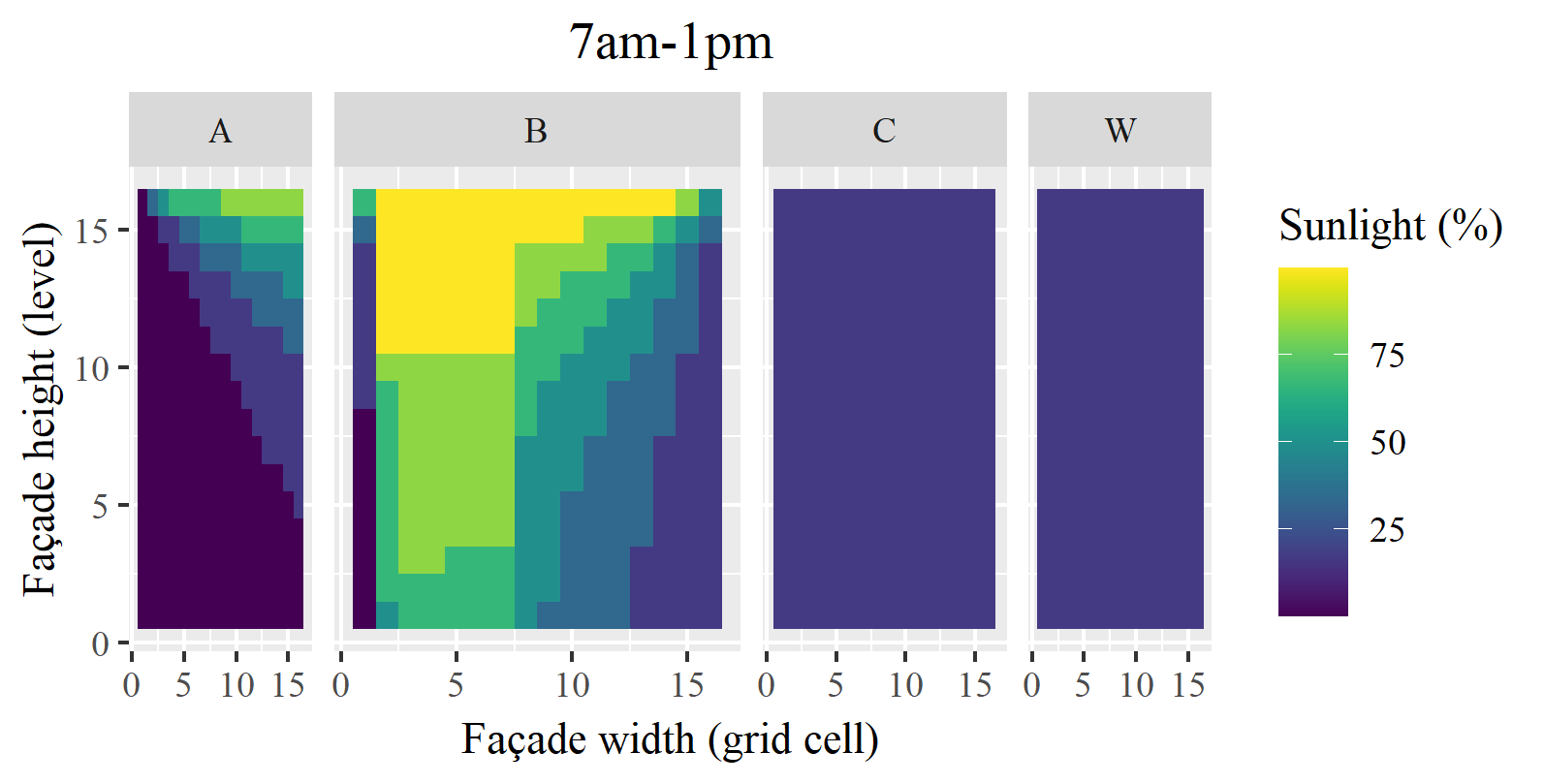}
\end{subfigure}

\begin{subfigure}{0.5\textwidth}
	\centering
	\includegraphics[width=0.9\linewidth]{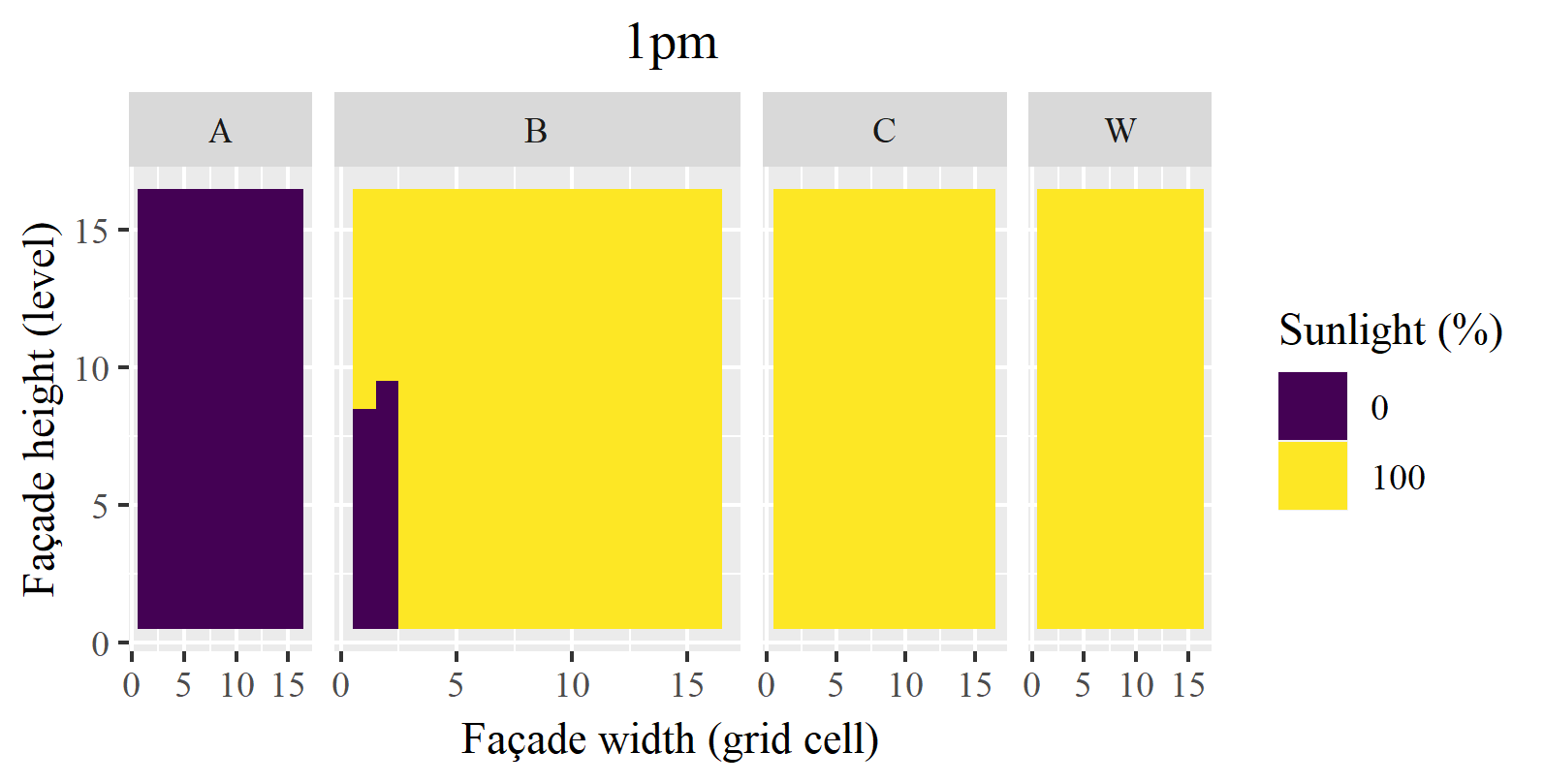} 
\end{subfigure}%
\begin{subfigure}{0.5\textwidth}
	\centering	
	\includegraphics[width=0.9\linewidth]{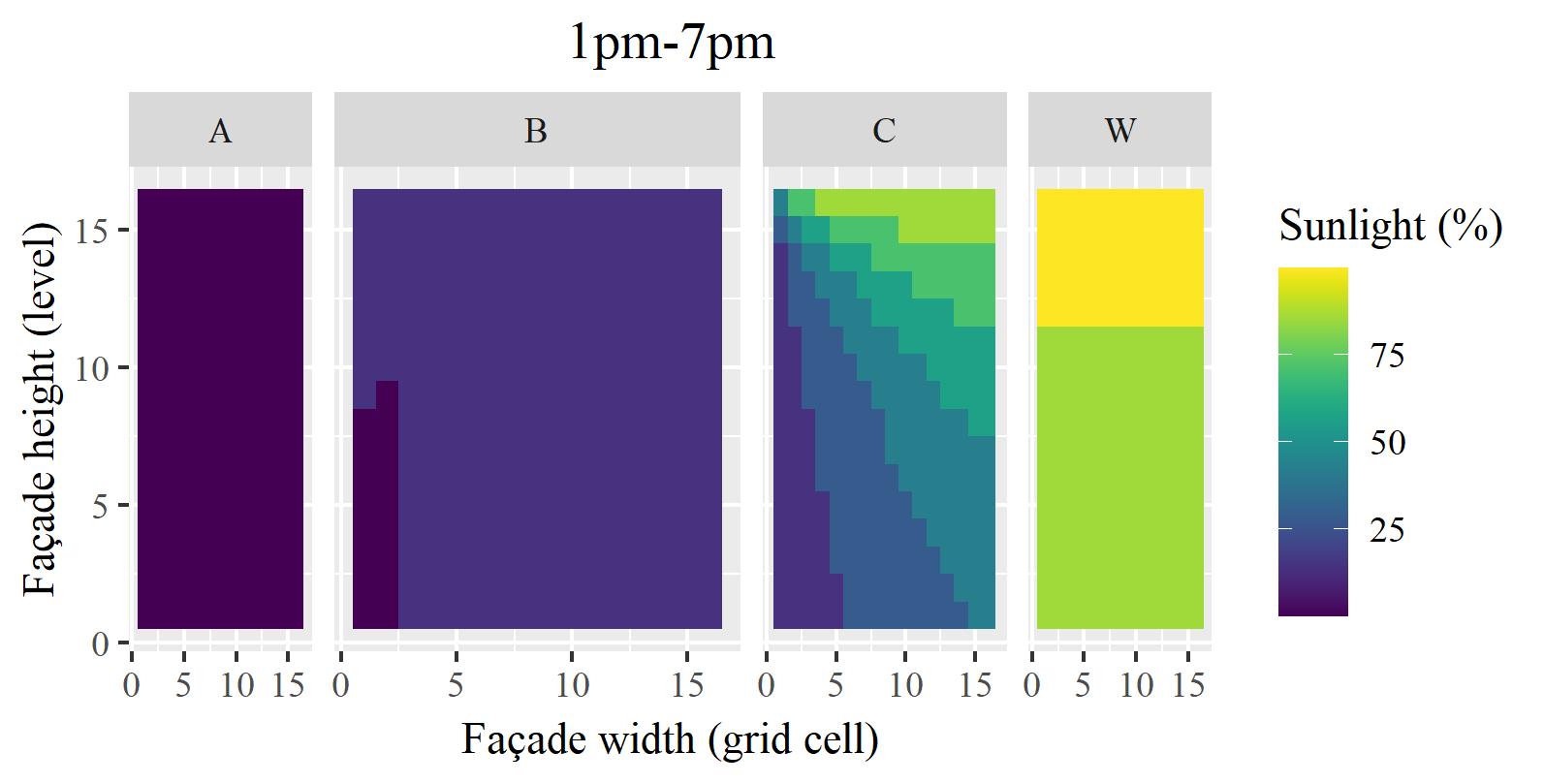}
\end{subfigure}

\begin{subfigure}{0.5\textwidth}
	\centering
	\includegraphics[width=0.9\linewidth]{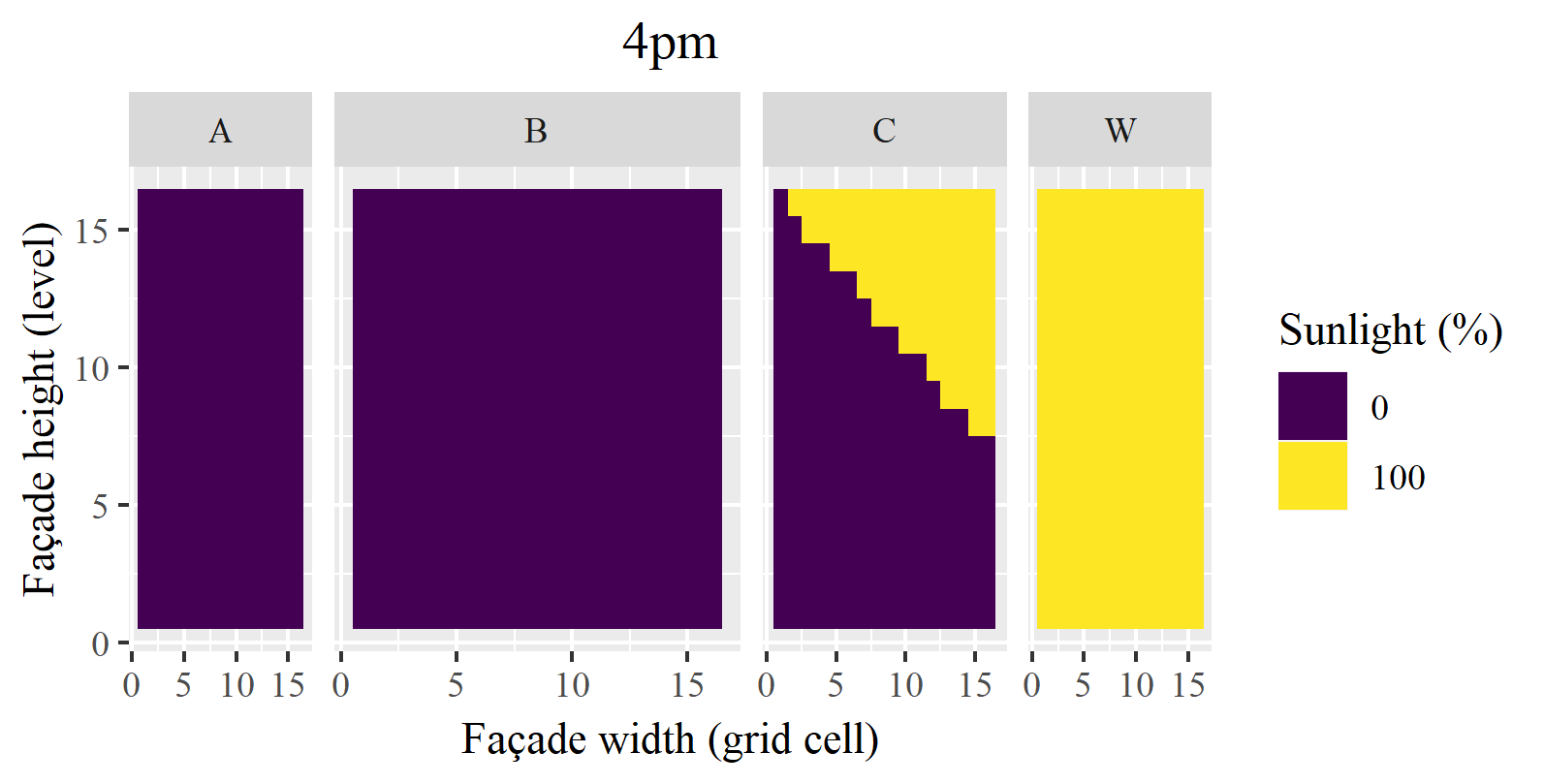} 
\end{subfigure}%
\begin{subfigure}{0.5\textwidth}
	\centering	
	\includegraphics[width=0.9\linewidth]{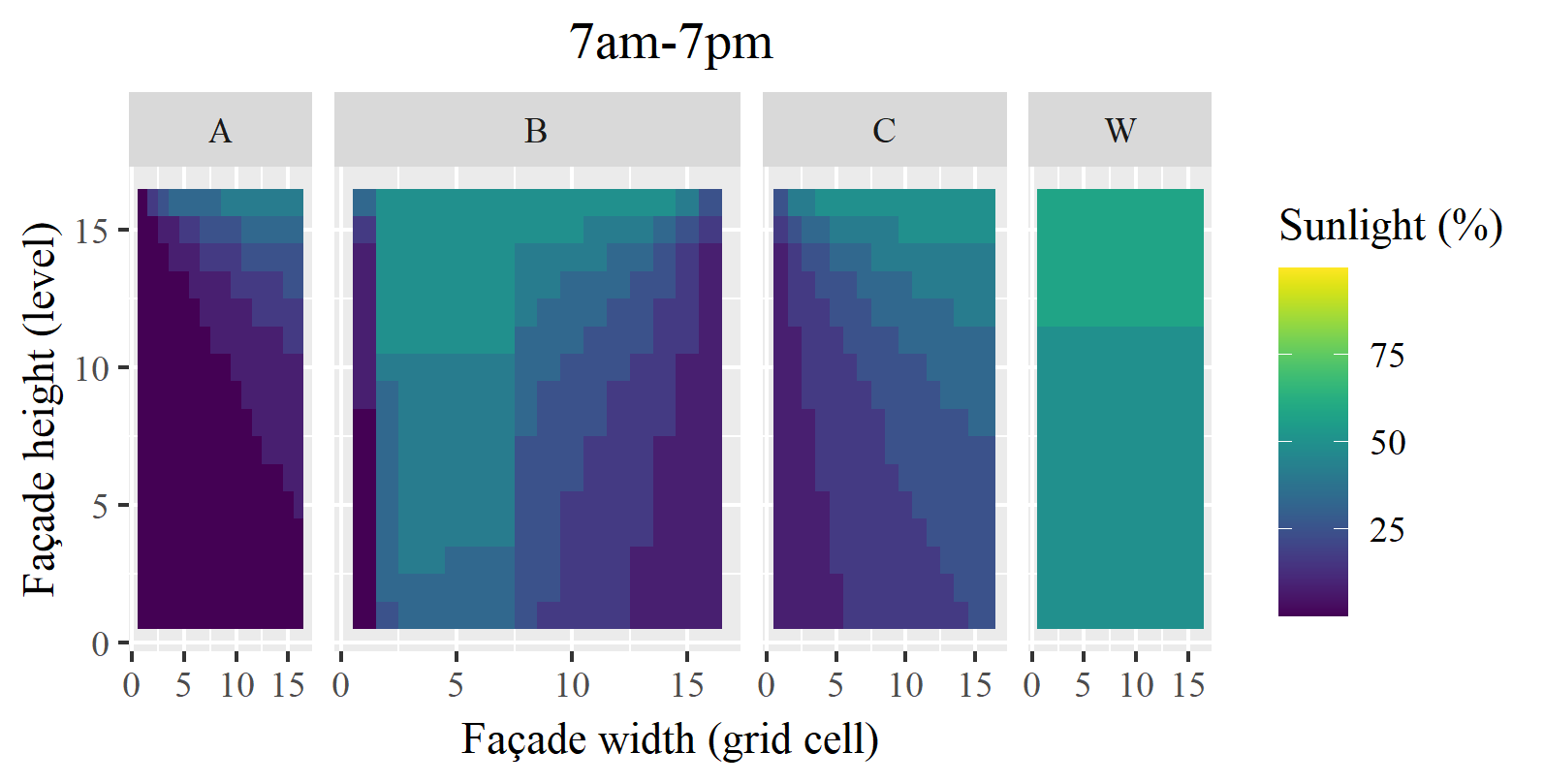}
\end{subfigure}
	
\caption{Shadow maps on a sunny day (17 Mar).}
\label{fig:shadowMaps}
\end{figure*}

Shadow maps (Figure \ref{fig:shadowMaps}) indicate that different areas of fa\c{c}ades A, B, C, and W receive direct sunlight for different durations on a sunny day. Fa\c{c}ade A received direct sunlight only for a small duration till noon which has been affected by shadowing effects of adjacent fa\c{c}ade and nearby building. The direct sunlight received on Fa\c{c}ade B from 7am--1pm and 1pm-7pm is affected by the shadowing effects of adjacent fa\c{c}ades (including fa\c{c}ade C) and buildings, and by the sun's diurnal motion respectively. Although fa\c{c}ades C and W face in the same direction and received direct sunlight for same duration from 7am--1pm, the percent of time they received direct sunlight from 1pm--7pm varied due to the shadowing effects of fa\c{c}ade B and nearby building respectively. It is also observed that the duration of time a fa\c{c}ade is exposed to direct sunlight increased with the height of the building. However, except for fa\c{c}ade W, there is only a small area of fa\c{c}ades B and C which received direct sunlight for more than equal to 50\% of the daytime (i.e. 7am--7pm).  While fa\c{c}ade B received most of this sunlight during the first half of the day, fa\c{c}ades C and W received it during the second half. Further, these areas are mostly located at level 11 \& above for fa\c{c}ade B and level 15 \& above for fa\c{c}ade C. Fa\c{c}ade A received direct sunlight for about 40\% of the daytime only on level 16. For rest of the daytime, all these fa\c{c}ades received indirect sunlight. 
As different crops have different light requirements, they may only achieve optimum growth when placed at certain fa\c{c}ades that can meet the crops' light requirements.

\subsubsection{PAR and DLI analysis based on lighting simulations}

\begin{figure*}[pos = htbp]

\begin{subfigure}{0.5\textwidth}
	\centering
	\includegraphics[width=0.9\linewidth]{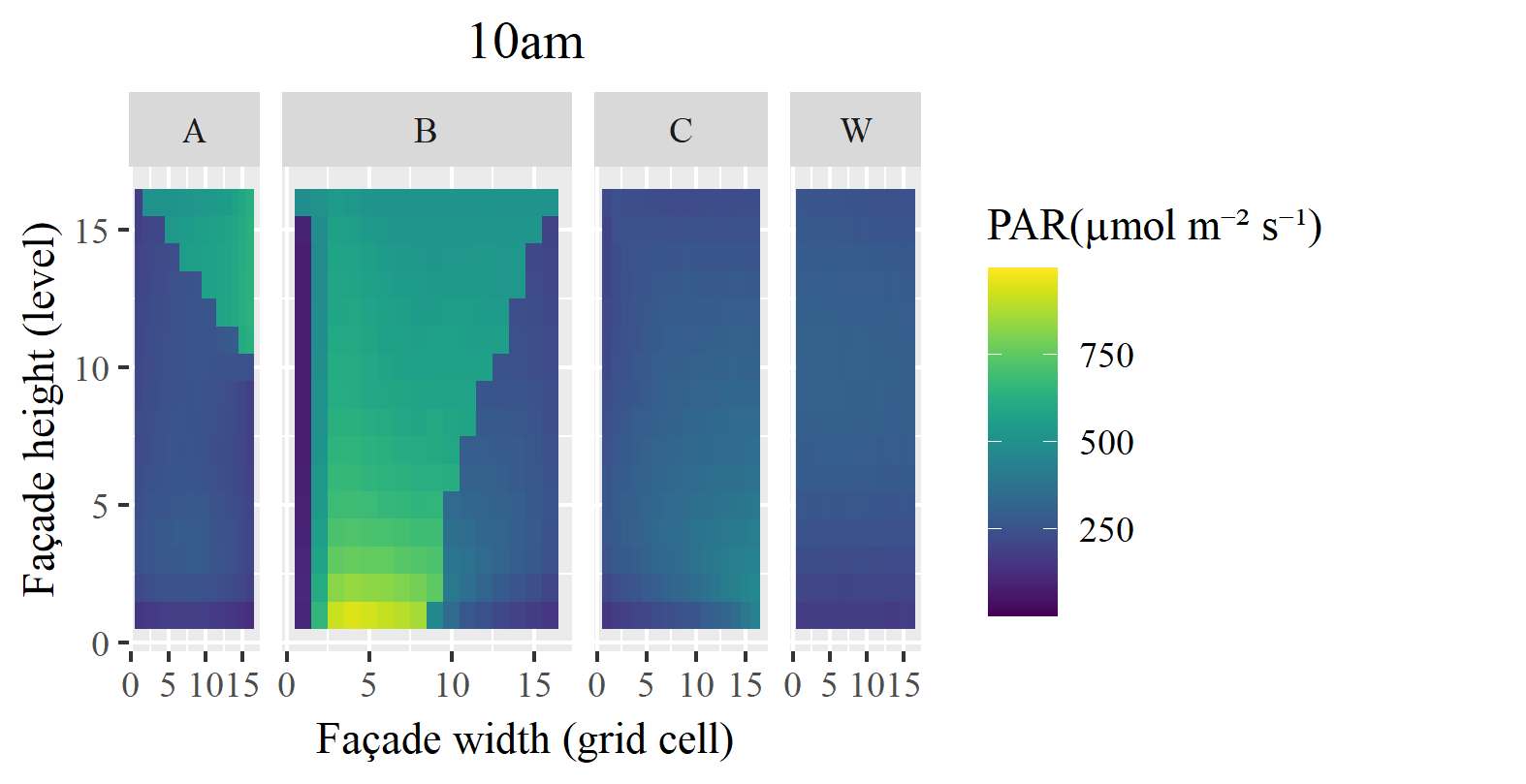} 
\end{subfigure}%
\begin{subfigure}{0.5\textwidth}
	\centering	
	\includegraphics[width=0.9\linewidth]{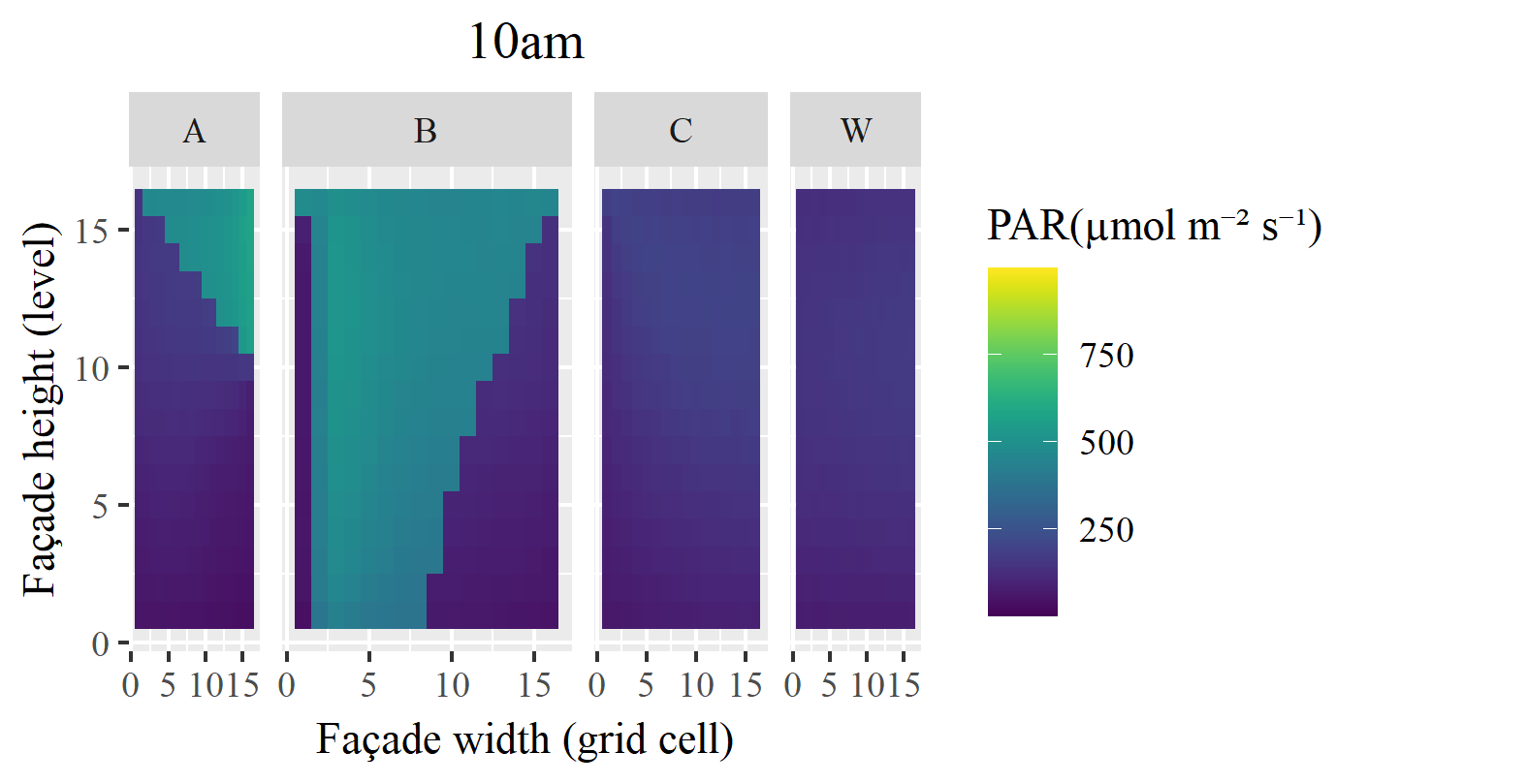}
\end{subfigure}

\begin{subfigure}{0.5\textwidth}
	\centering
	\includegraphics[width=0.9\linewidth]{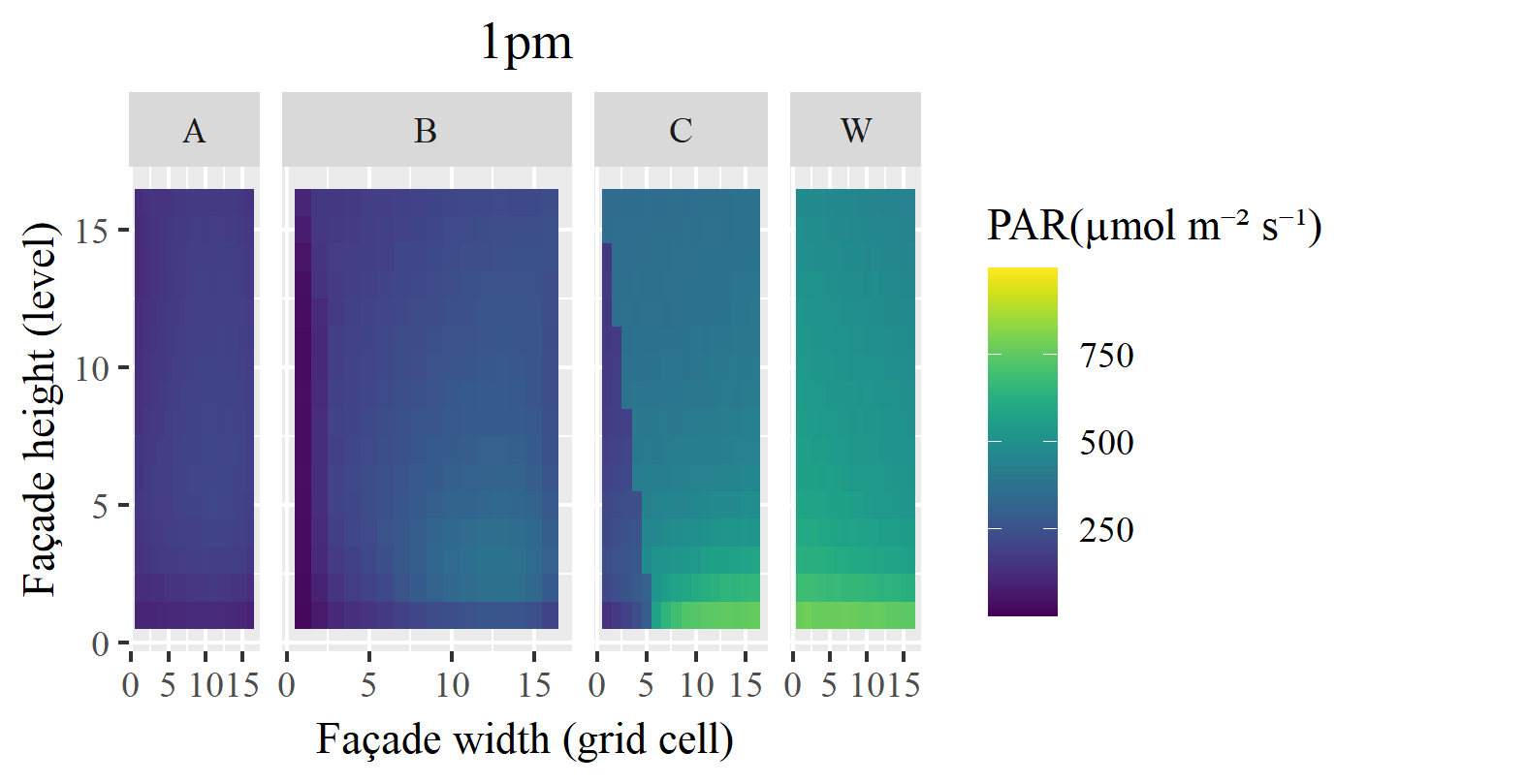} 
\end{subfigure}%
\begin{subfigure}{0.5\textwidth}
	\centering	
	\includegraphics[width=0.9\linewidth]{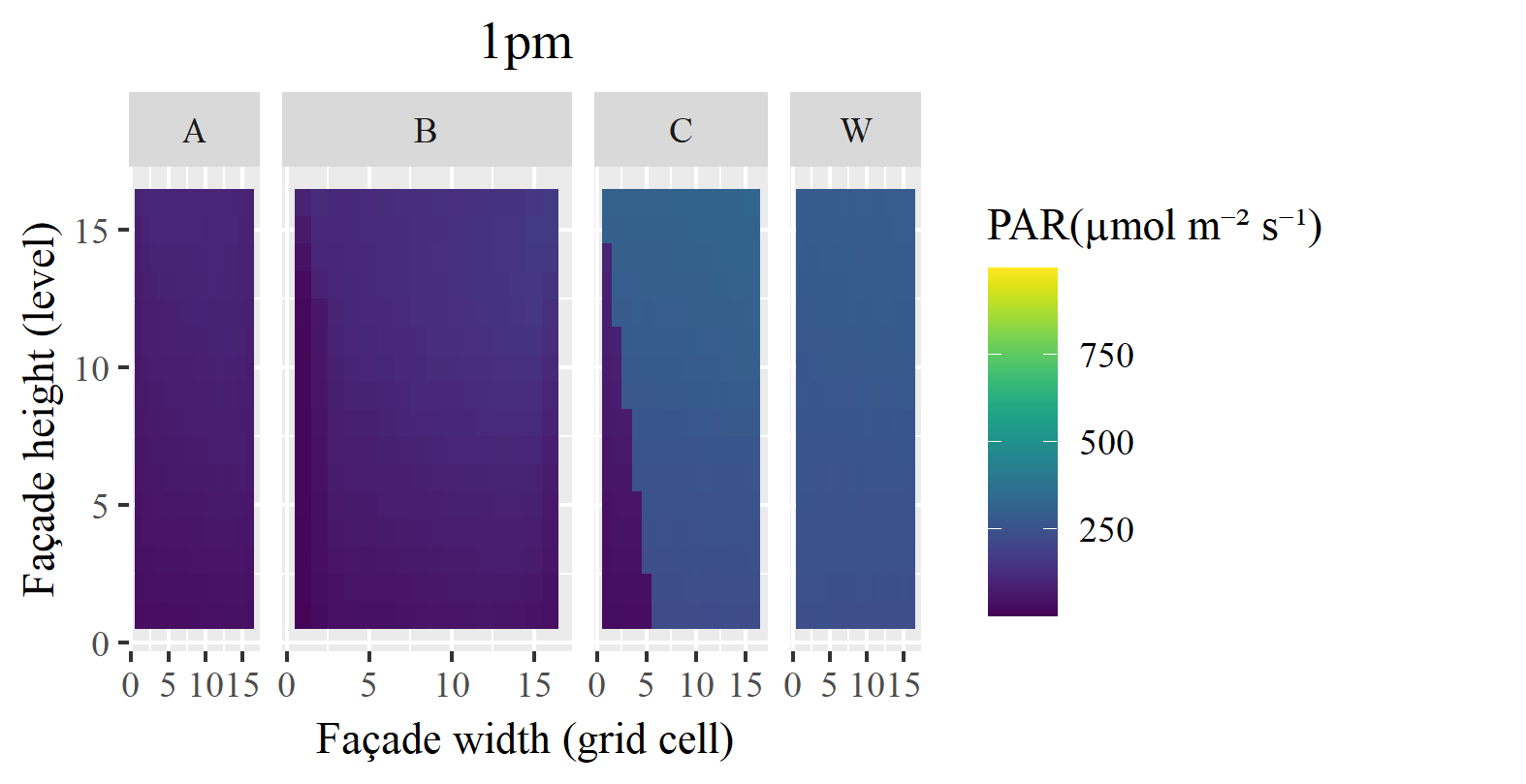}
\end{subfigure}

\begin{subfigure}{0.5\textwidth}
	\centering
	\includegraphics[width=0.9\linewidth]{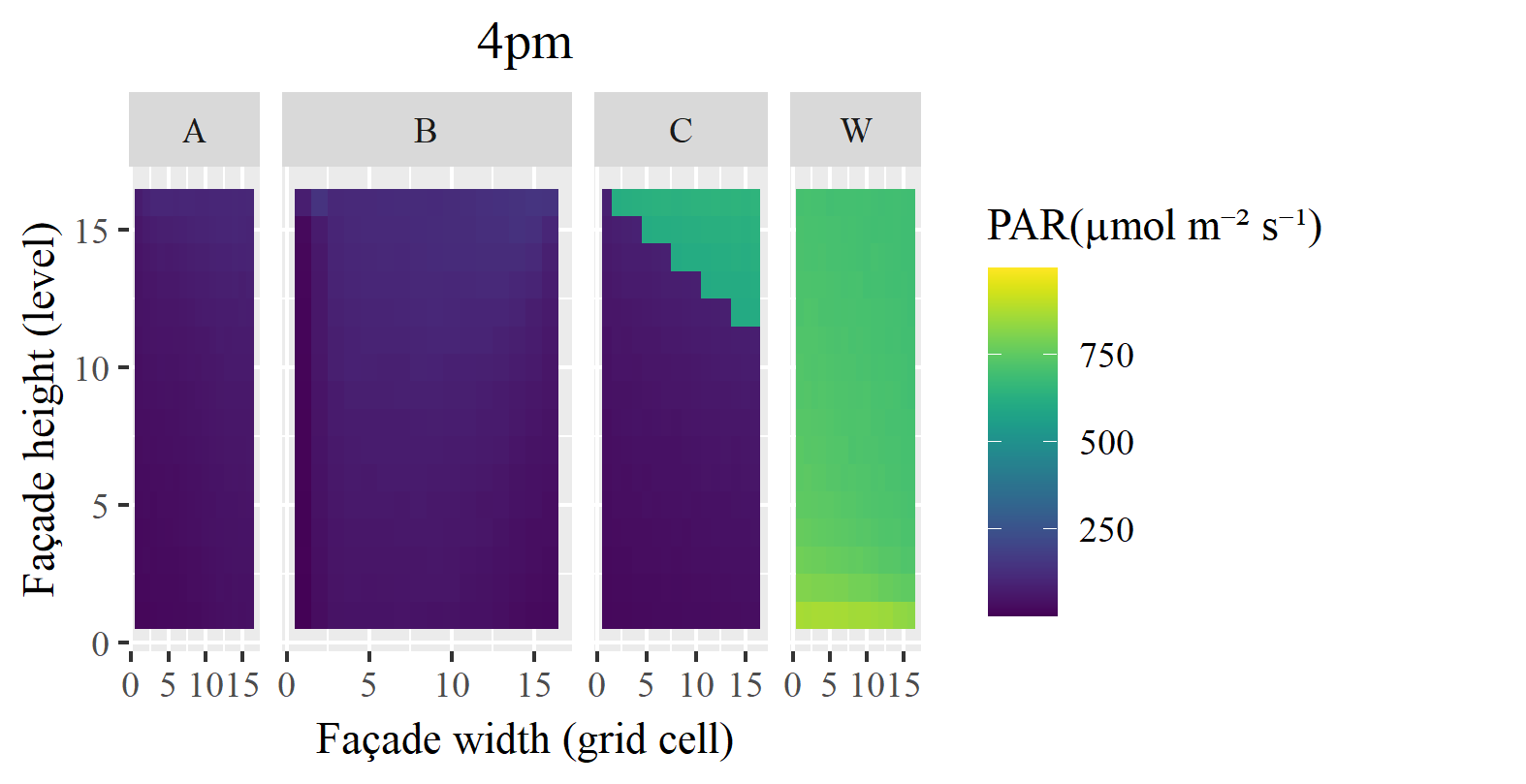} 
	\caption{With ground reflections.}
	\label{fig:Ground}
\end{subfigure}%
\begin{subfigure}{0.5\textwidth}
	\centering	
	\includegraphics[width=0.9\linewidth]{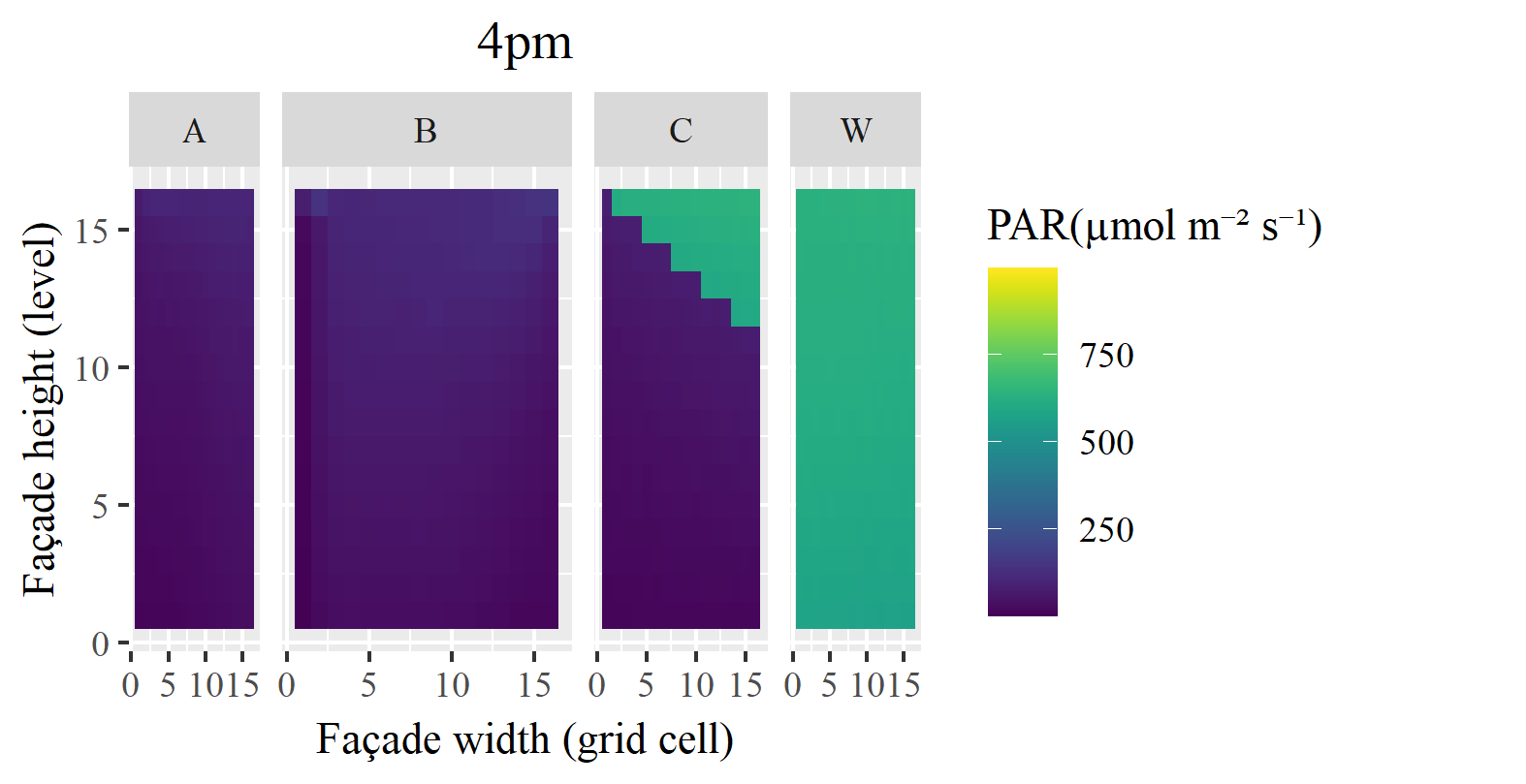}
	\caption{Without ground reflections.}
	\label{fig:noGround}
\end{subfigure}
	
\caption{PAR distribution on fa\c{c}ades on a sunny day (17 Mar).}
\label{fig:SunnyPARsim}
\end{figure*}

Figure \ref{fig:Ground} shows that on a sunny day PAR exceeds 500$\Psi$ on fa\c{c}ades A and B in the morning hour except for locations which were shadowed by adjacent fa\c{c}ade(s) and building.  However, as the day progresses, PAR tend to remain below 300$\Psi$. A trend reversal was observed in case of fa\c{c}ades C and W where PAR at or below 300$\Psi$ was observed in the morning followed by PAR exceeding 500$\Psi$ in the later hours of the day. Further, enormously high PAR observed on the lower levels of fa\c{c}ades B, C, and W were due to ground reflections (Figure \ref{fig:noGround}) which are a component of indirect sunlight. It is admitted here that such PAR levels from ground reflections are not observed in practice, can be attributed to high reflectance given to ground plane \citep[p.171]{ludu2013}, and is a limitation of the simulations carried out in VI-Suite. In the absence of ground reflections, PAR was found to increase with height on these fa\c{c}ades during these times. PAR on the same level of the fa\c{c}ade was found to be uniform except in cases where shadows were casted by adjacent fa\c{c}ade(s) and building. In such cases, on an average, PAR was reduced by 58\% of the PAR observed in unshaded conditions on the same level (Table \ref{tbl3} in Appendix \ref{sec:appendix}).

\begin{figure*}[pos = htbp]

\begin{subfigure}{0.5\textwidth}
	\centering
	\includegraphics[width=0.9\linewidth]{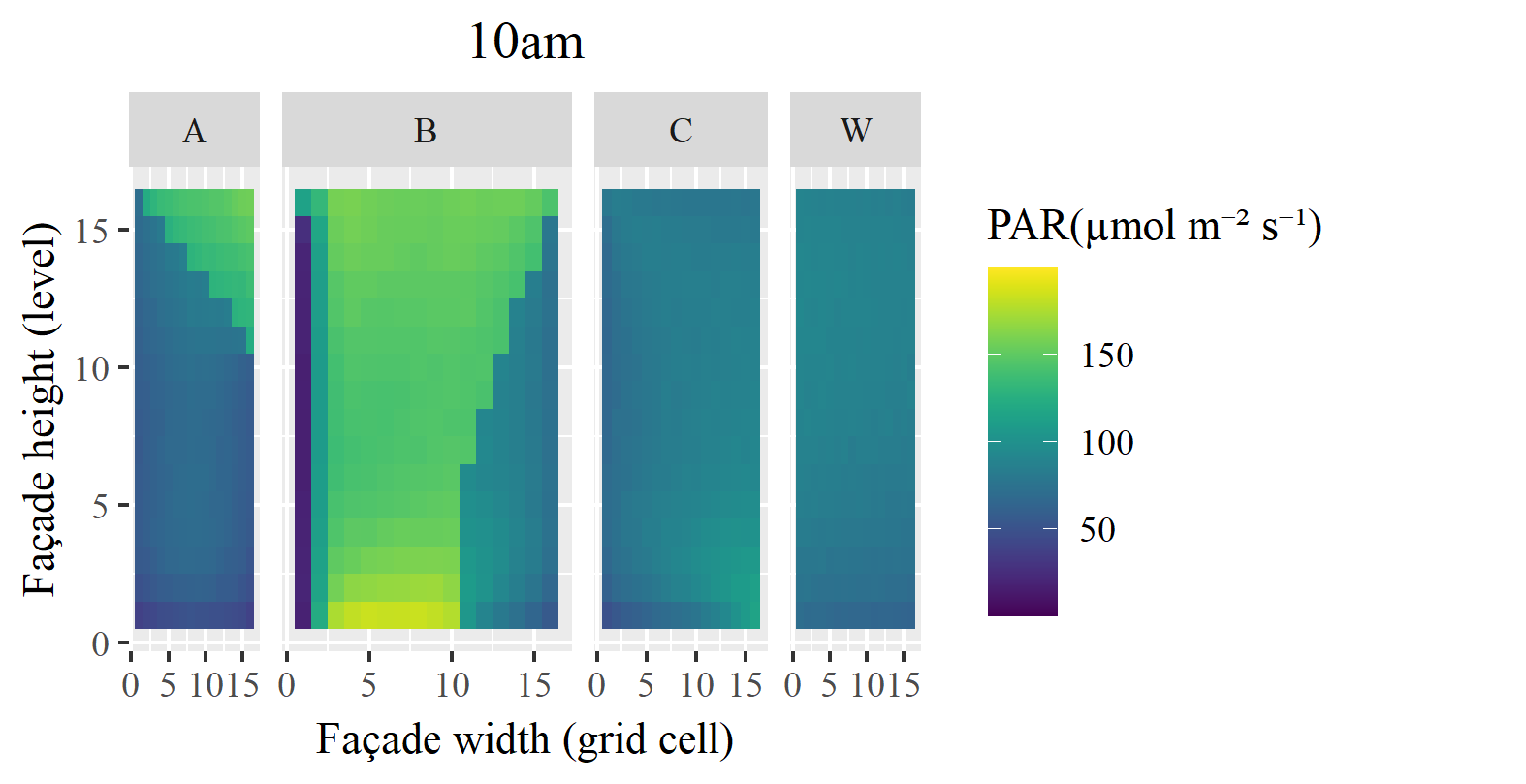} 
\end{subfigure}%
\begin{subfigure}{0.5\textwidth}
	\centering	
	\includegraphics[width=0.9\linewidth]{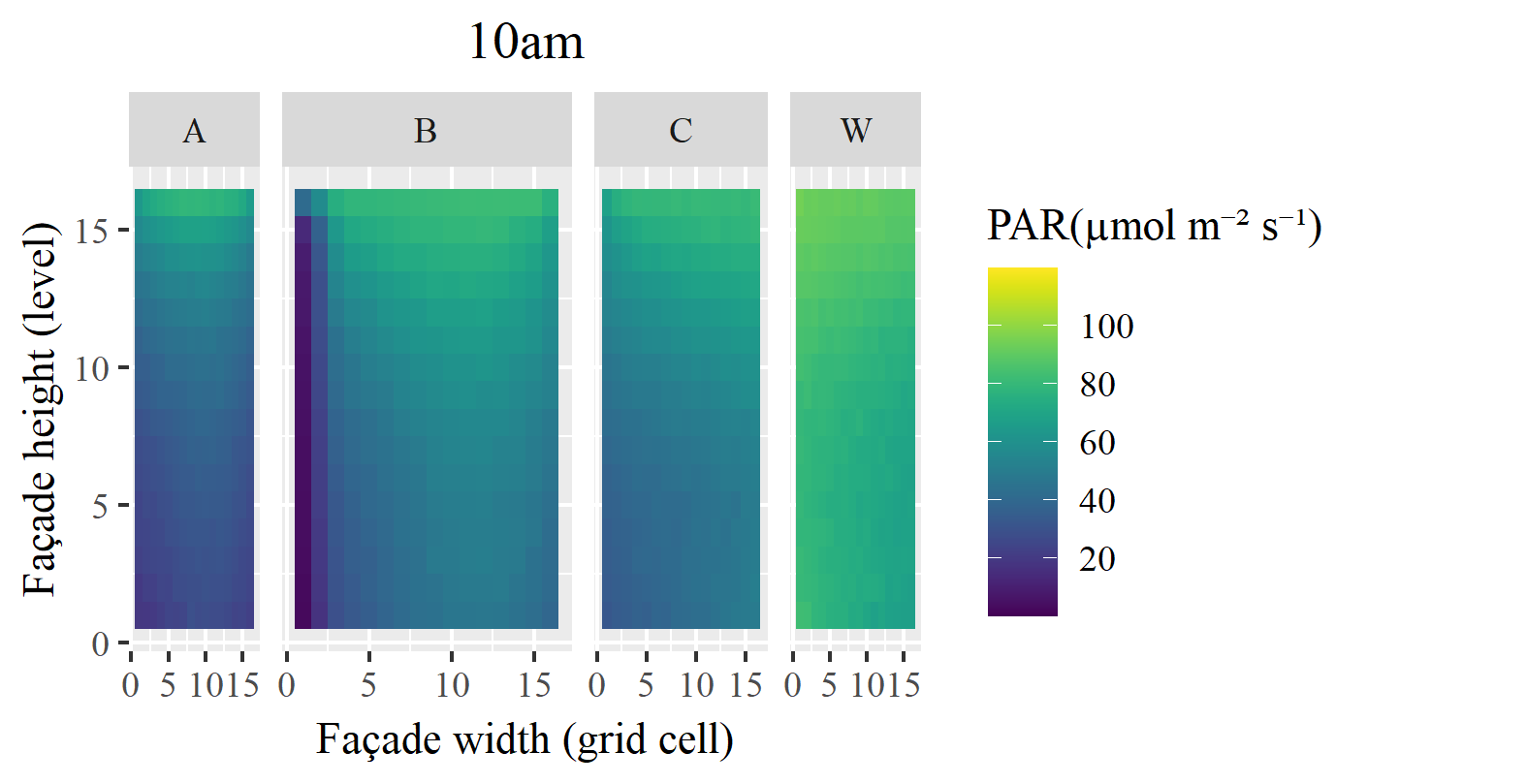}
\end{subfigure}

\begin{subfigure}{0.5\textwidth}
	\centering
	\includegraphics[width=0.9\linewidth]{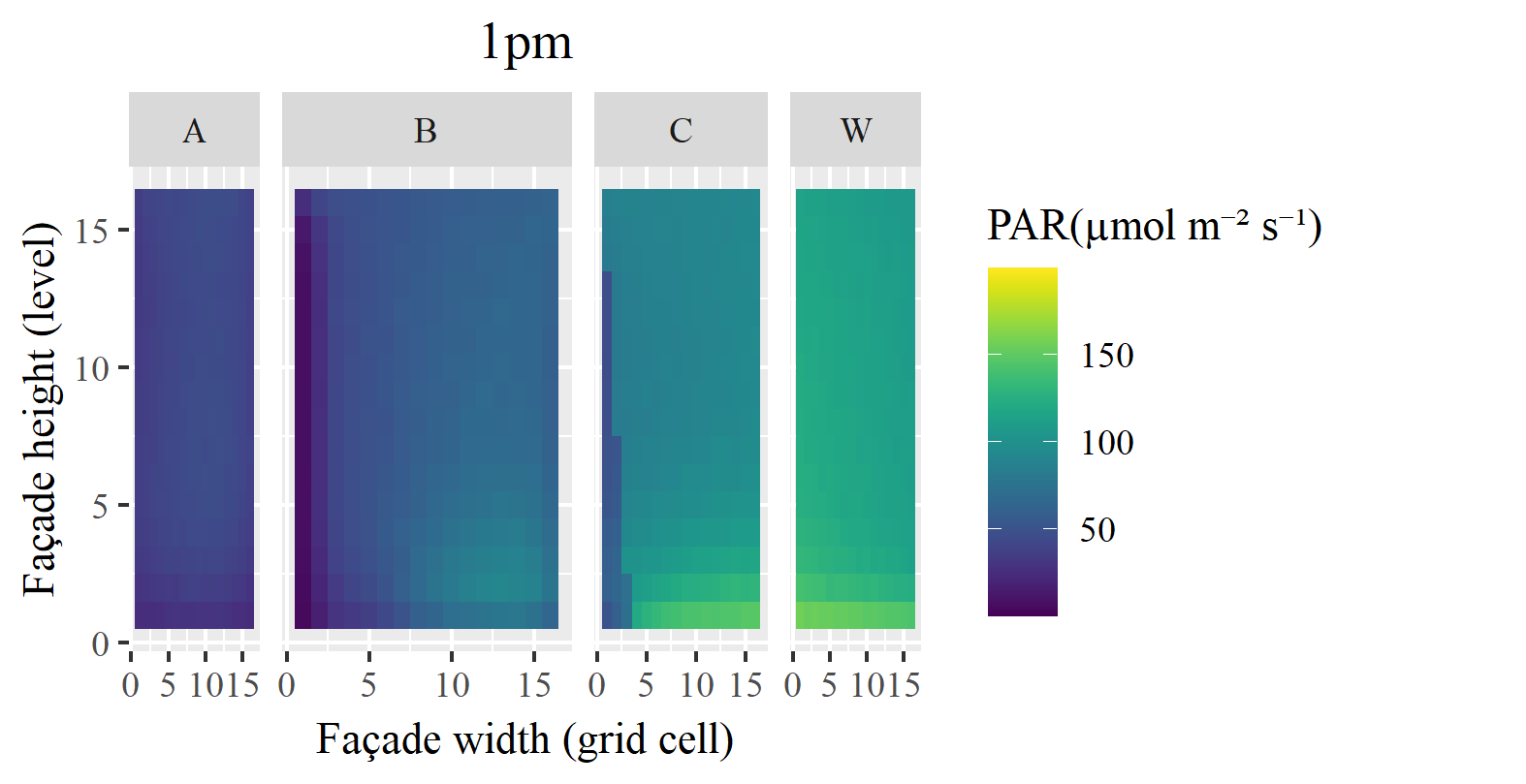} 
\end{subfigure}%
\begin{subfigure}{0.5\textwidth}
	\centering	
	\includegraphics[width=0.9\linewidth]{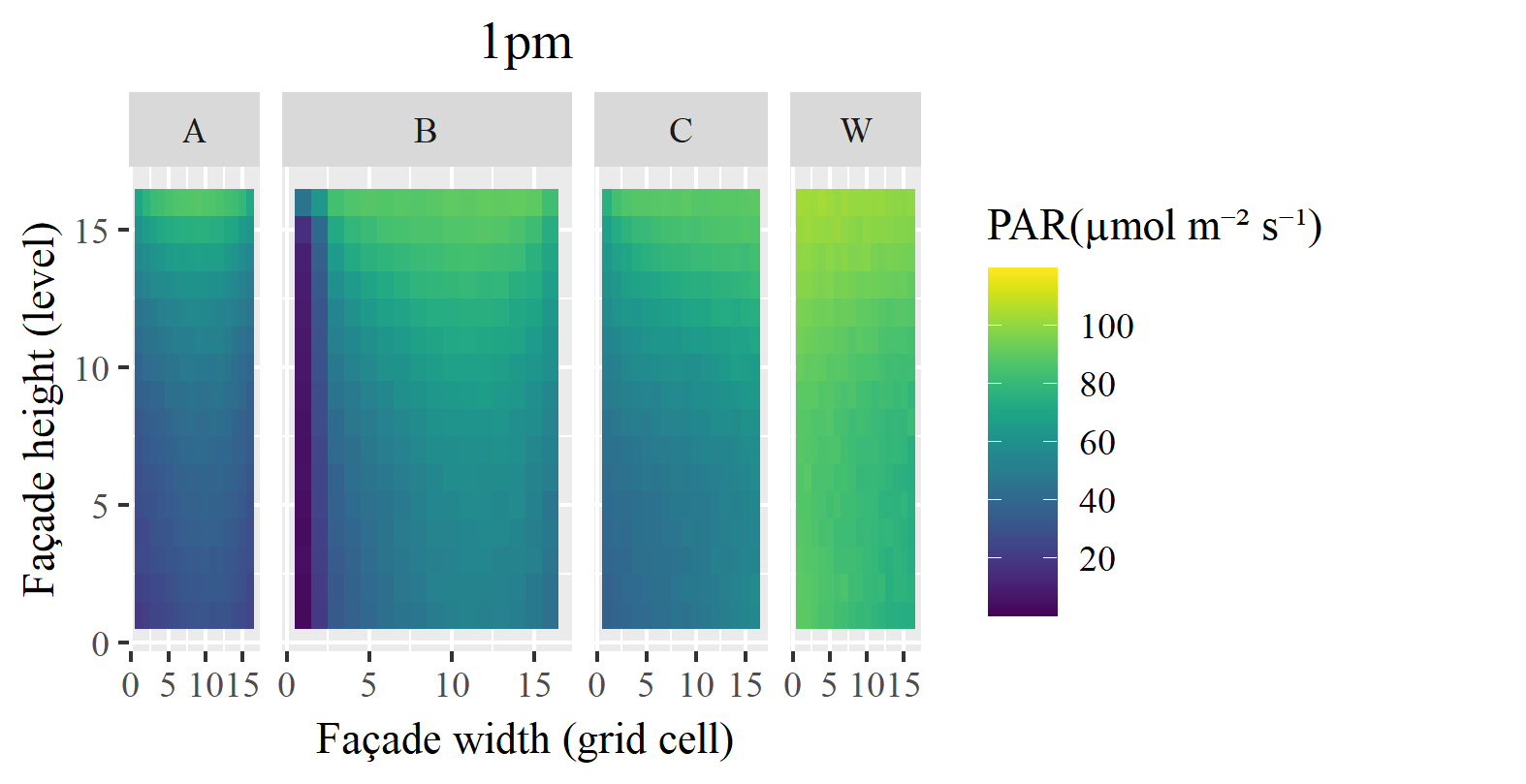}
\end{subfigure}

\begin{subfigure}{0.5\textwidth}
	\centering
	\includegraphics[width=0.9\linewidth]{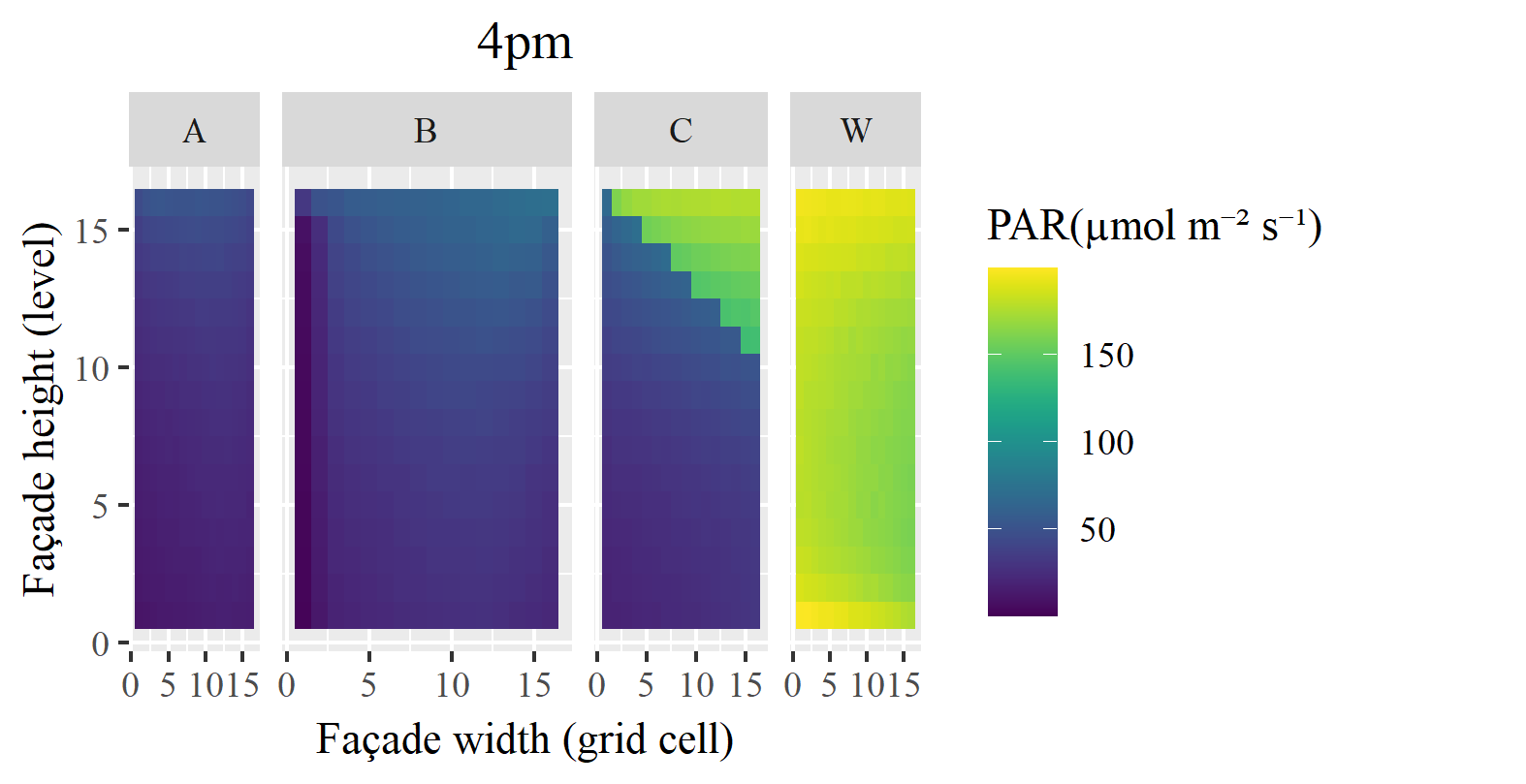} 
	\caption{Partly cloudy day (09 Mar).}
	\label{fig:partlyCloudy}
\end{subfigure}%
\begin{subfigure}{0.5\textwidth}
	\centering	
	\includegraphics[width=0.9\linewidth]{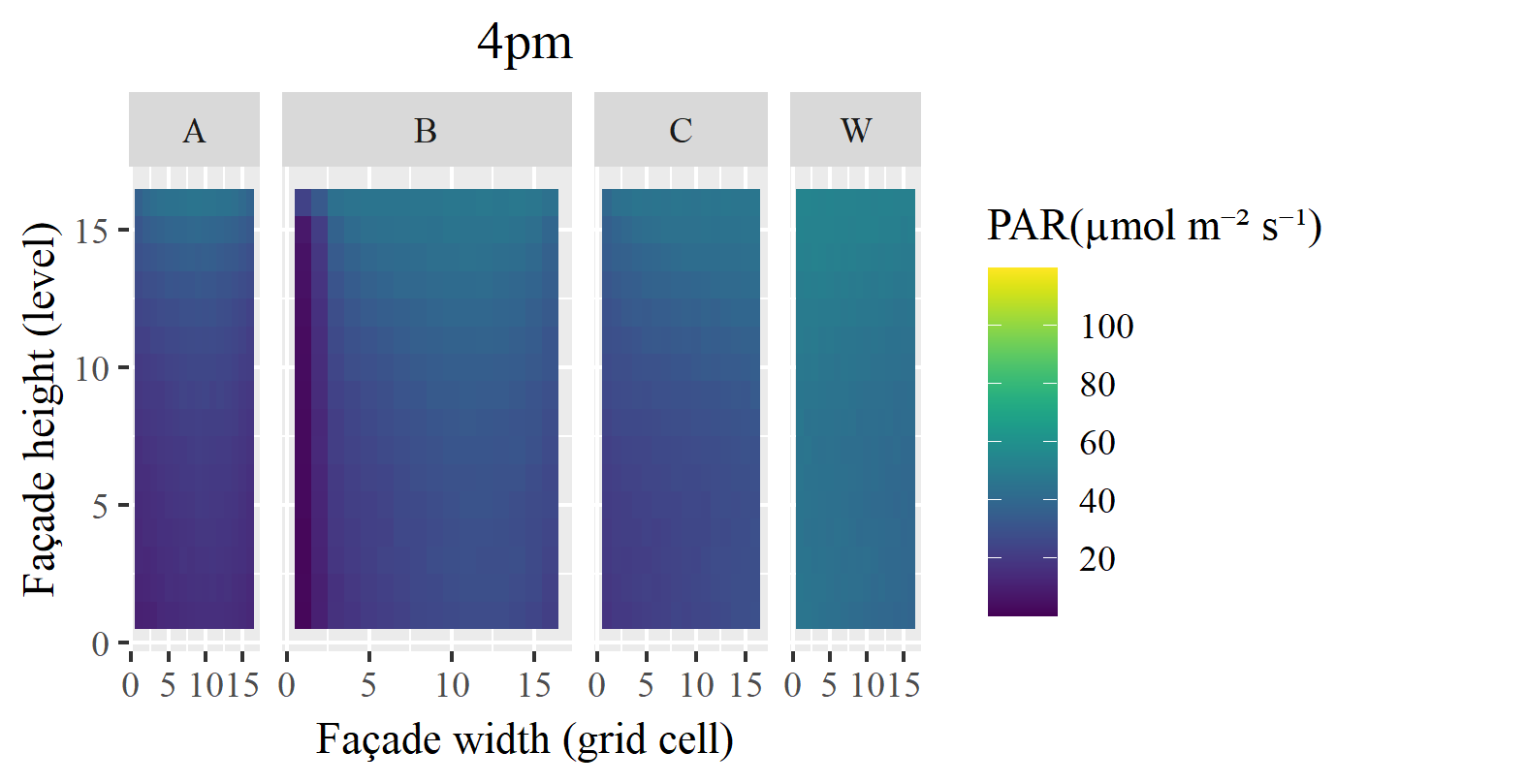}
	\caption{Cloudy day (03 Apr).}
	\label{fig:Cloudy}
\end{subfigure}
	
\caption{PAR distribution on fa\c{c}ades on a partly cloudy and cloudy day.}
\label{fig:CloudyPARsim}
\end{figure*}

PAR on different fa\c{c}ades during a partly cloudy day (Figure \ref{fig:partlyCloudy}) followed similar trend as that of a sunny day that is, PAR decreased on fa\c{c}ades A and B and increased on fa\c{c}ades C and W through the day. While the maximum PAR observed on fa\c{c}ades A and B was about 150$\Psi$ and 200$\Psi$ respectively pre-noon, it remained below 50$\Psi$ and 100$\Psi$ respectively post-noon. PAR on fa\c{c}ades C and W at 10am, 1pm, and 4pm remained below 120$\Psi$, 150$\Psi$, and 200$\Psi$ respectively. PAR showed an increasing trend with height at 10am for fa\c{c}ades A and B and for all fa\c{c}ades at 4pm. In other cases, PAR remained in a narrow range with no incremental trend with height. On the same level of a fa\c{c}ade, on an average, PAR under shaded conditions was reduced by 40\% of the PAR in unshaded conditions (Table \ref{tbl3} in Appendix \ref{sec:appendix}). Abnormally high PAR on lower levels of fa\c{c}ades were also observed on partly cloudy day.

Shadowing effects did not have a significant impact on PAR distribution on a cloudy day (Figure \ref{fig:Cloudy}) primarily due to the cloud cover and no particular trend in PAR on fa\c{c}ades was observed through the day. Further, maximum PAR observed for all the fa\c{c}ades at 10am, 1pm, and 4pm was around 90$\Psi$, 100$\Psi$, and 50$\Psi$ respectively. For all the fa\c{c}ades, PAR was found to increase with height while remaining in a narrow range.

\begin{figure}[pos = htbp]
	\centering
	\includegraphics[width=0.5\textwidth]{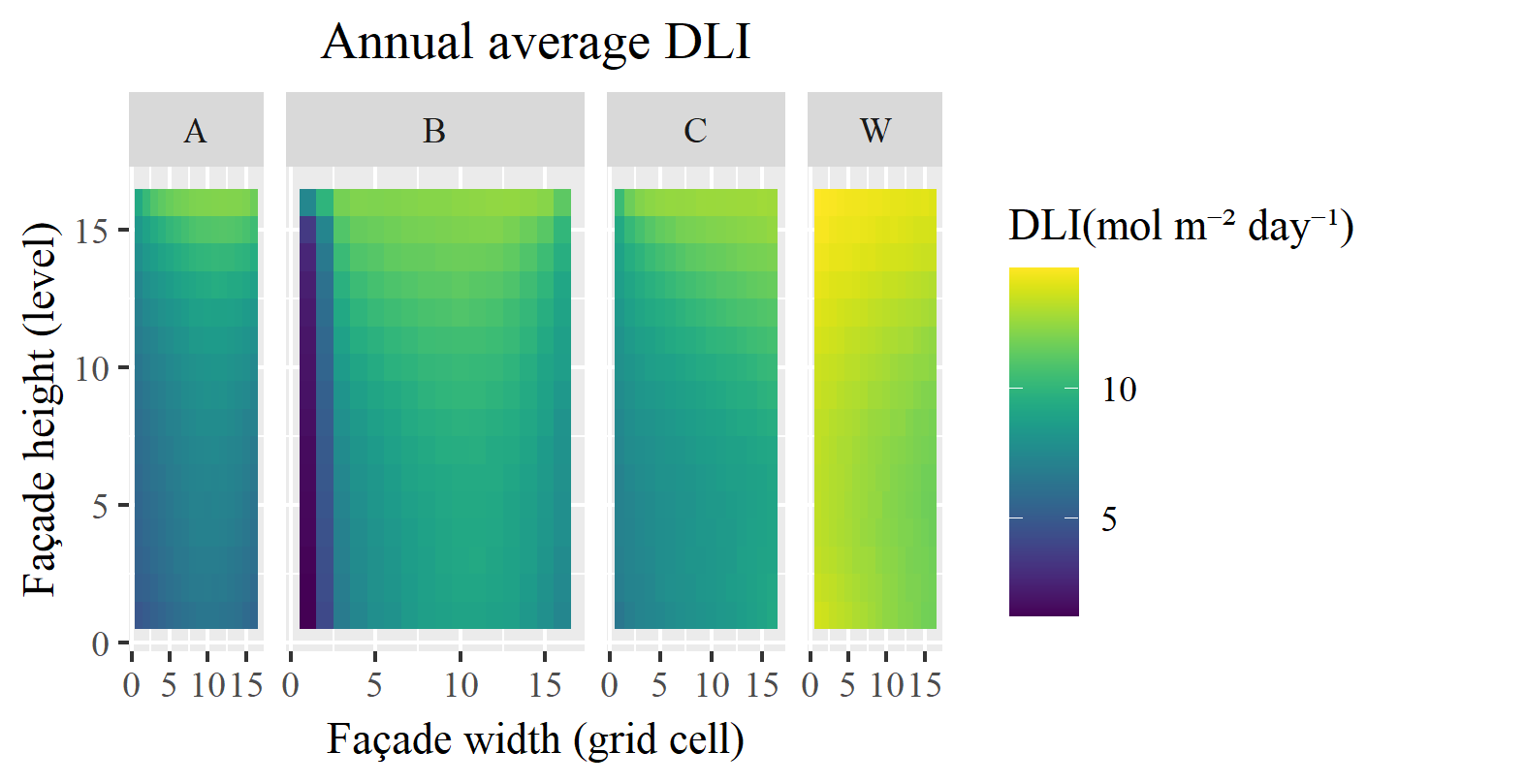}
	\caption{\label{fig:AnnualavgDLI}Annual average DLI distribution on fa\c{c}ades.}
\end{figure}

The annual average DLI on fa\c{c}ades A, B, C, and W ranged between 5--12 $\Phi$, 1--12 $\Phi$, 7--13 $\Phi$, and 12--15 $\Phi$ respectively (Figure \ref{fig:AnnualavgDLI}). For all the fa\c{c}ades, these values increased with height. Further, average DLI above 9$\Phi$ was observed in all grid cells of fa\c{c}ade W and in some grid cells of fa\c{c}ades A, B, and C above levels 12, 6, and 7 respectively. The number of grid cells with these values on fa\c{c}ades A, B, and C also increased at higher levels. The variation in average DLI on a given level, if any was the cumulative result of the shadowing effects of the adjacent fa\c{c}ade(s) and building(s) and changing weather conditions as per the EPW data.   

\begin{figure*}[pos = htbp]

\begin{subfigure}{0.5\textwidth}
	\centering
	\includegraphics[width=0.9\linewidth]{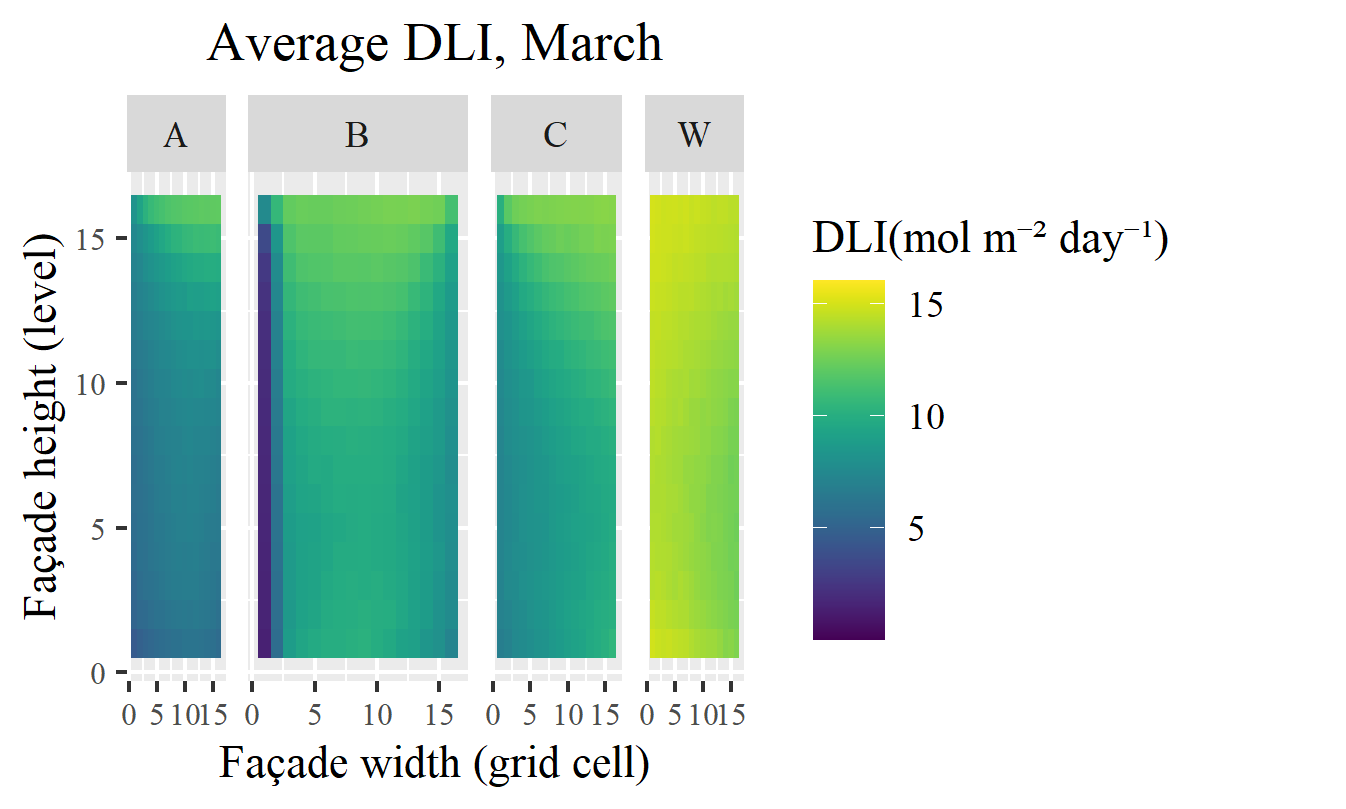} 
\end{subfigure}%
\begin{subfigure}{0.5\textwidth}
	\centering	
	\includegraphics[width=0.9\linewidth]{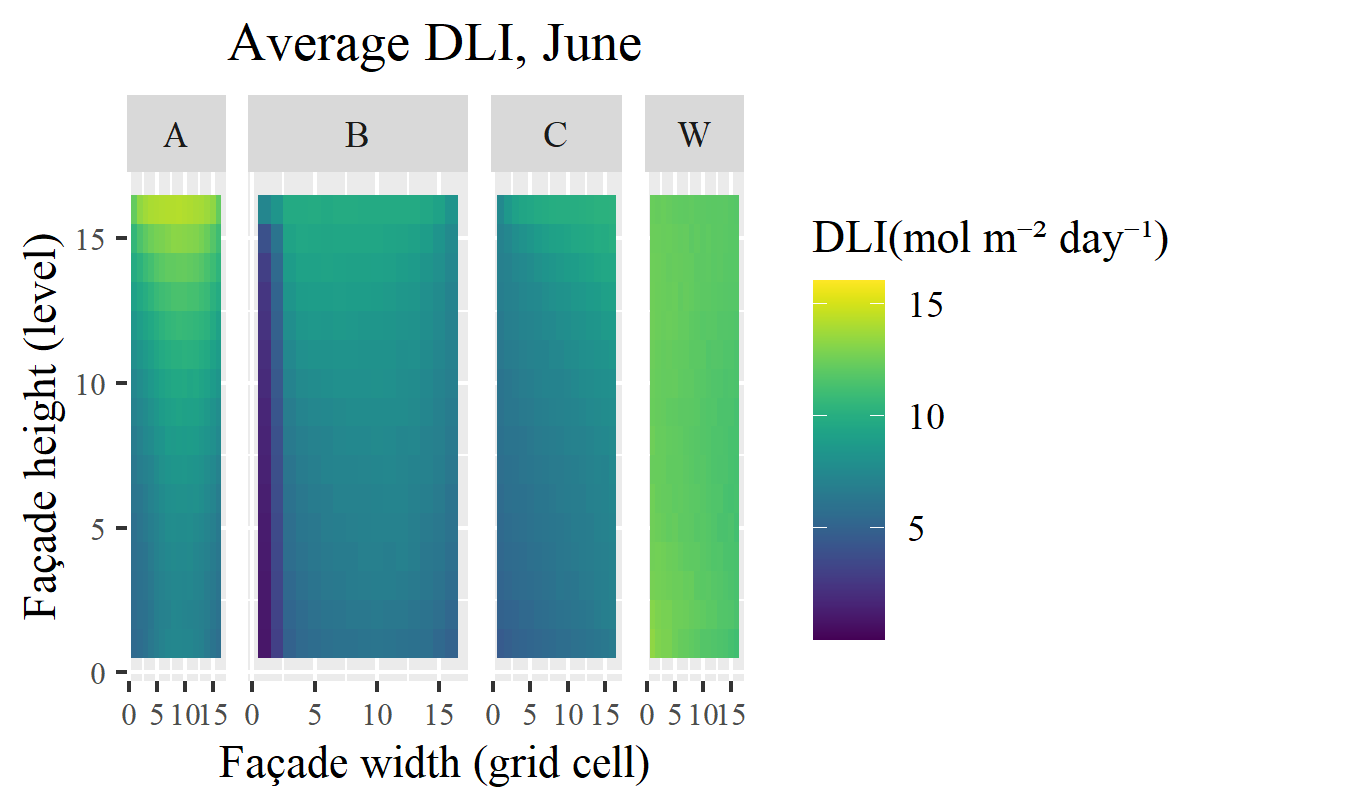}
\end{subfigure}

\begin{subfigure}{0.5\textwidth}
	\centering
	\includegraphics[width=0.9\linewidth]{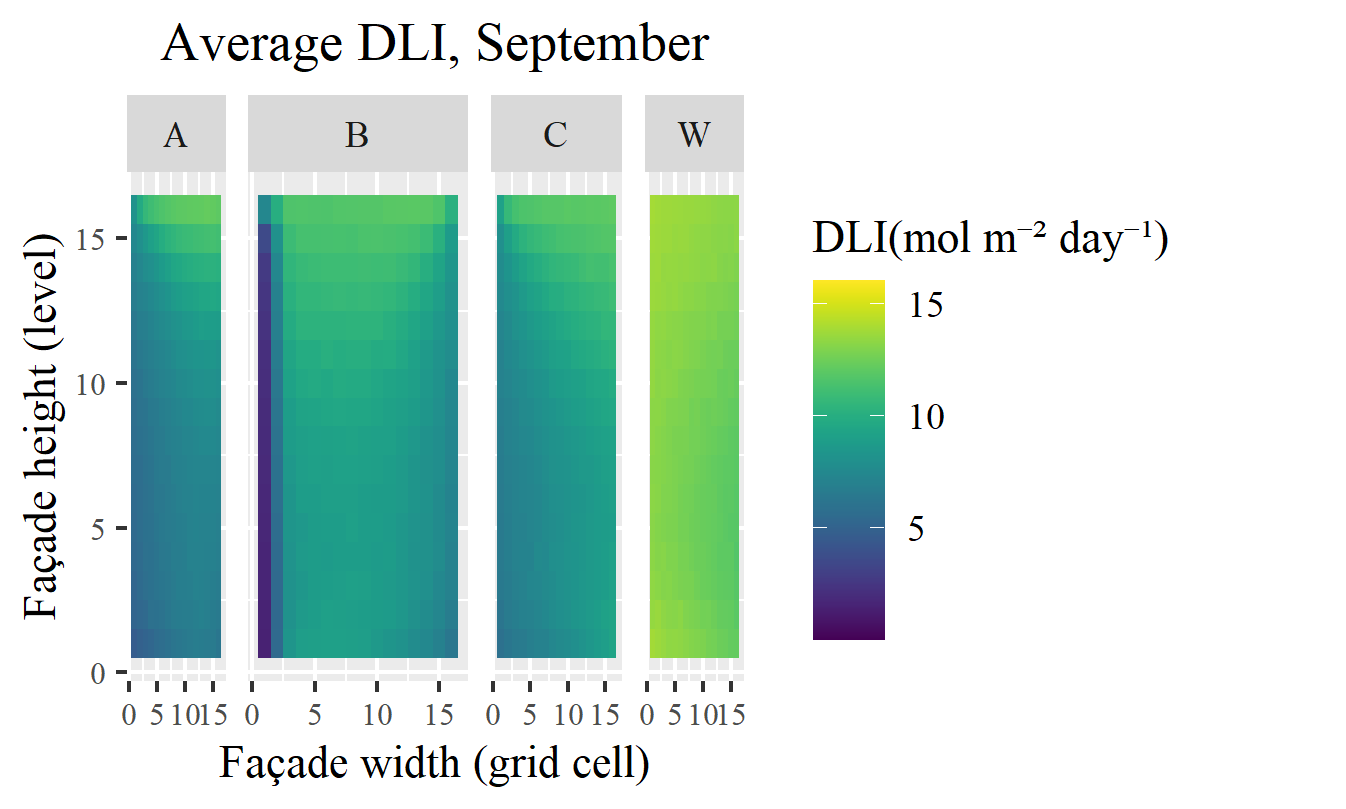} 
\end{subfigure}%
\begin{subfigure}{0.5\textwidth}
	\centering	
	\includegraphics[width=0.9\linewidth]{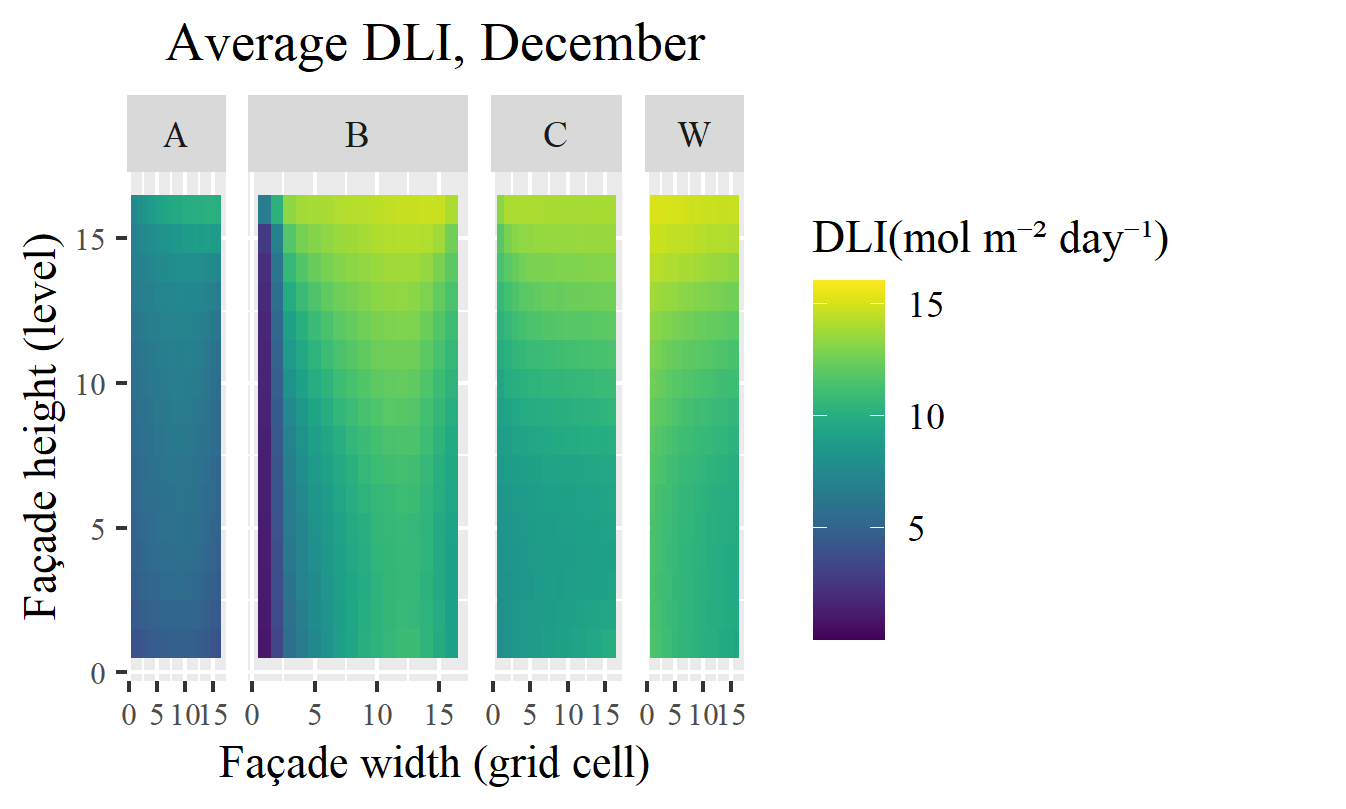}
\end{subfigure}
	
\caption{Average DLI distribution on fa\c{c}ades for selected months.}
\label{fig:MonthlyavgDLI}
\end{figure*}

The monthly average DLI for March, June, September, and December (Figure \ref{fig:MonthlyavgDLI}) demonstrated similar trends as that of annual average DLI. While the range of monthly average DLI for the fa\c{c}ades remained almost same, the levels at which the average DLI exceeded 9$\Phi$ in the grid cells varied significantly across the months for all the fa\c{c}ades except fa\c{c}ade W (Table \ref{tbl4} in Appendix \ref{sec:appendix}). 

\subsubsection{Comparison of measured and simulated PAR}

\begin{figure*}[pos = htbp]

\begin{subfigure}{\textwidth}
	\centering
	\includegraphics[width=0.5\linewidth]{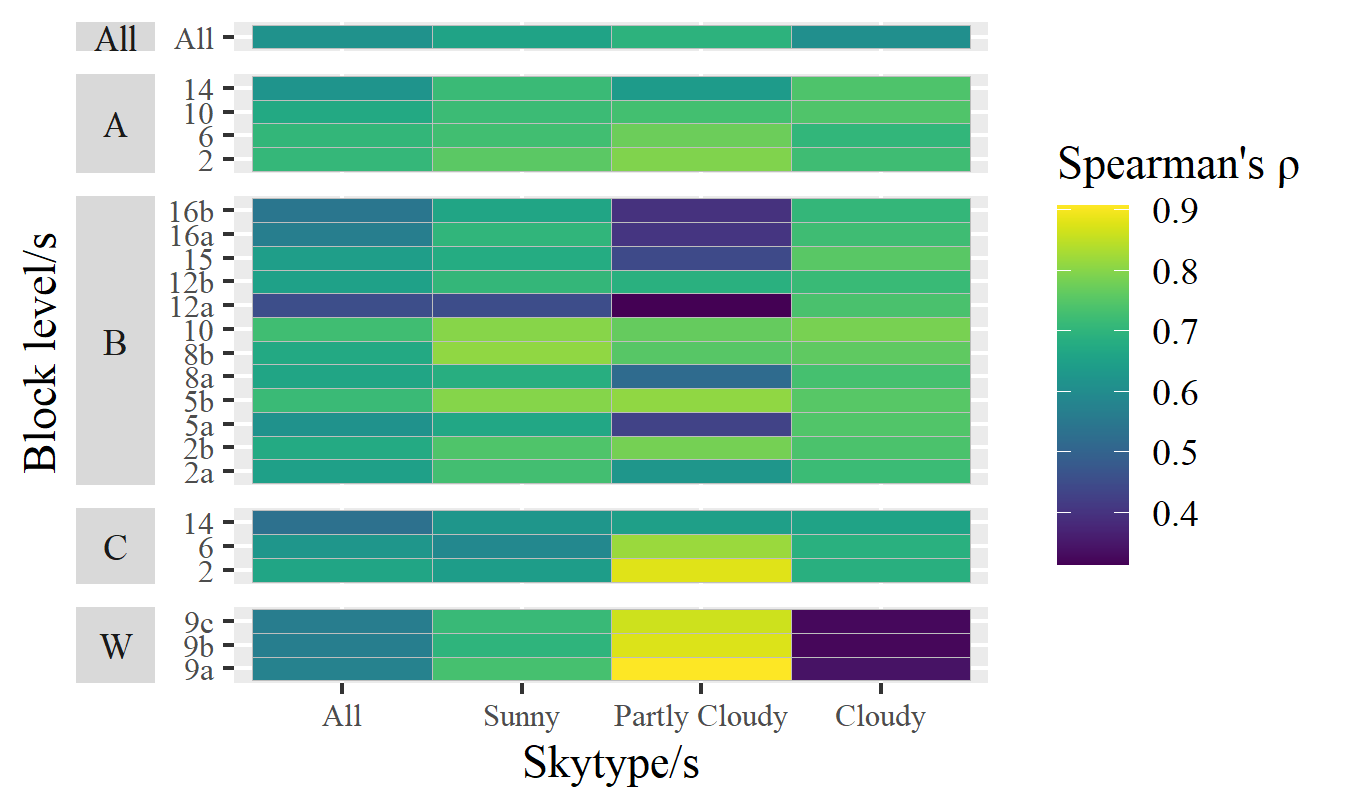} 
\end{subfigure}

\begin{subfigure}{0.5\textwidth}
	\centering
	\includegraphics[width=0.9\linewidth]{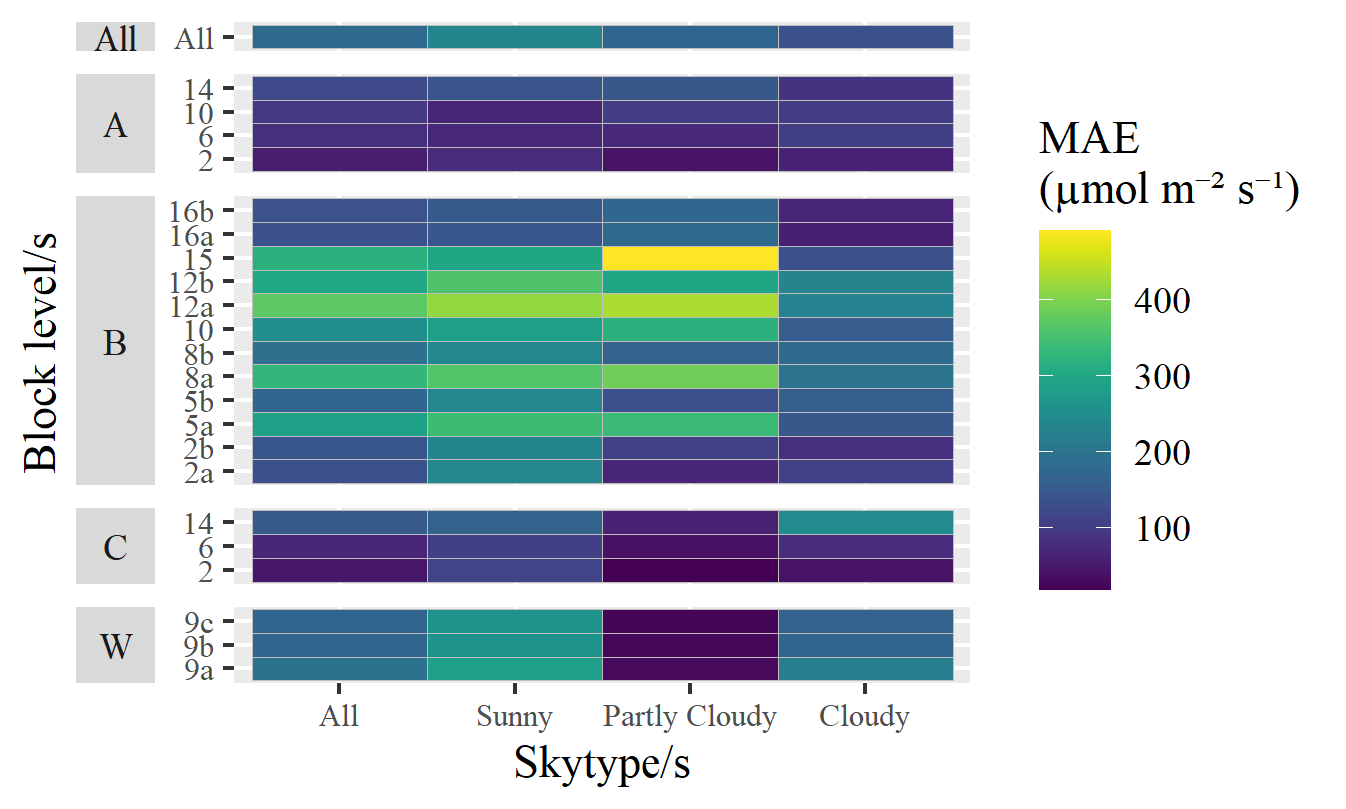} 
\end{subfigure}%
\begin{subfigure}{0.5\textwidth}
	\centering	
	\includegraphics[width=0.9\linewidth]{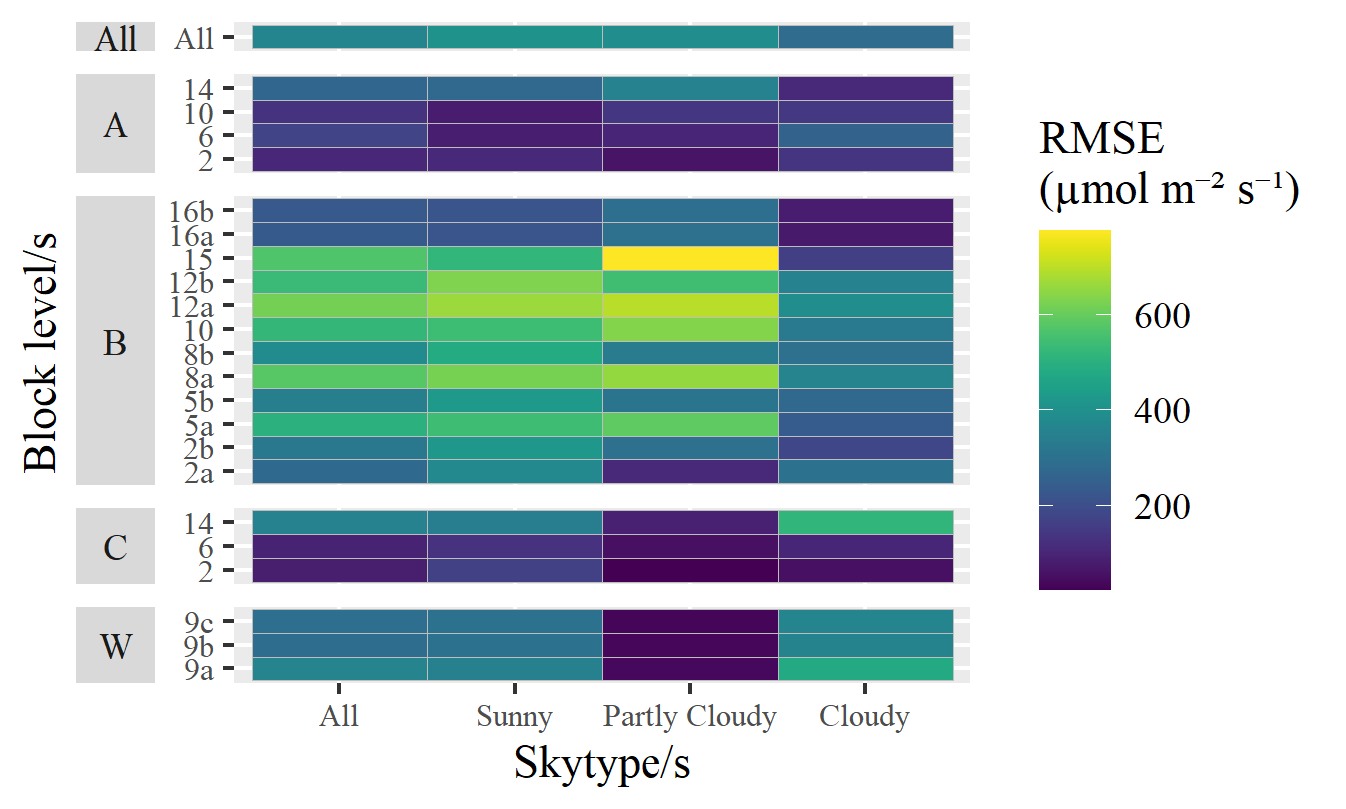}
\end{subfigure}
	
\caption{Metrics for comparing measured and simulated PAR.}
\label{fig:Metrics}
\end{figure*}

Spearman's $\rho$ between measured and simulated PAR, considering all skytypes and sensor locations on fa\c{c}ades together (Figure \ref{fig:Metrics}), was determined to be 0.62. MAE and RMSE were found to be 176.9$\Psi$ and 365.2$\Psi$ respectively. Location wise analysis showed that under all skytypes, $\rho$ was between 0.52 and 0.74 for all sensor locations except for 12a\footnote{Sensors placed on the same level of a fa\c{c}ade were labelled as `a', `b', and so on starting from the left in Figure \ref{fig:sensor}.}  where it was 0.45. In addition, relatively high values of MAE and RMSE were observed for sensors placed on fa\c{c}ade B at levels 8, 12, and 15. 

Under sunny skytype, all sensor locations together exhibited $\rho$, MAE, and RMSE of 0.66, 233.2$\Psi$, and 407.3$\Psi$ respectively. Observations similar to all skytypes were also made for sunny skytype for individual sensor locations with $\rho$ between 0.58 and 0.82 except for 12a ($\rho$ = 0.45) and relatively high values of MAE and RMSE for sensors at levels 8 and 12 of fa\c{c}ade B.
 
When all sensor locations were considered together, $\rho$, MAE, and RMSE under partly cloudy skytype were comparable to sunny and all skytypes with values of 0.70, 170.1$\Psi$, and 389.7$\Psi$ respectively. However, location wise analysis showed that although MAE and RMSE were relatively high for sensors at levels 8, 10, 12, and 15 on fa\c{c}ade B, $\rho$ was between 0.51 and 0.92 except for sensors at 5a, 12a, 15, 16a, and 16b. 
 
Under cloudy skytype, $\rho$, MAE, and RMSE were 0.61, 135.9$\Psi$, and 287.5$\Psi$ respectively for all sensor locations considered together. $\rho$ varied between 0.65 and 0.80 except for sensors at fa\c{c}ade W where it was about 0.33. MAE and RMSE remained relatively low except for sensor at level 14 of fa\c{c}ade C.

\subsection{Discussion}
Sun path analysis, shadow map analysis, and spatio-temporal analysis of PAR and DLI suggest that PAR and DLI at a location in the building are dependent on the building's shape and orientation, the sun's diurnal and annual motion, skytypes, and shadowing effects of the building's own fa\c{c}ades and nearby buildings. Further, in contrast to \cite{rubioetal2016}, it is not possible to adjudge the significance of one factor over another as all these factors have a cumulative effect on PAR and DLI at a given location and their individual significance may change with change in study area and study period.

In agreement with \cite{songetal2018}, PAR on the fa\c{c}ades in the study area remained largely similar as the fa\c{c}ades were exposed to direct sunlight for similar durations owing to the sun's diurnal motion. While fa\c{c}ades A and B experienced higher PAR during the former half of the day, fa\c{c}ades C and W experienced it during the latter half. However, this pattern was visible only for sunny and partly cloudy skytypes. No such pattern was observed under cloudy skytype primarily due to cloud cover. Further, PAR on the fa\c{c}ades was significantly reduced moving from sunny to partly cloudy to cloudy skytypes.
 
In addition to sunny skytype \citep{songetal2018}, PAR on these fa\c{c}ades increased with height for cloudy skytype as well. However, no such trend was discerned for partly cloudy skytype as it is characterized by bouts of sunlight and cloud cover through the day. In consonance with the findings of \cite{tanismail2014}, on the same level of the fa\c{c}ade, PAR was mainly affected due to shadowing effects of adjacent fa\c{c}ade(s) and/or building. Based on some test cases, PAR reduction due to shadowing effects was found to be more under sunny skytype as compared to partly cloudy skytype.       

The annual as well as monthly average DLI ranged from 1 to 15 $\Phi$ at different locations on these fa\c{c}ades with largely similar values on fa\c{c}ades across months. Confirming the finding of \cite{songetal2018}, average DLI increased with height. However, in real world, this trend as well as the higher average DLI observed on some fa\c{c}ades may be affected due to presence of trees and other built structures at lower levels. Except for fa\c{c}ade W, locations with average DLI exceeding 9$\Phi$ were mainly located at higher levels of the fa\c{c}ades. This may be attributed to the self-shadowing and shadowing effects of adjacent buildings on the lower levels \citep{rubioetal2016}. The dry phase of Northeast Monsoon during January--March \citep{climatend} and the sun's position in the southern hemisphere leading to higher PAR on the fa\c{c}ades can be the reasons for large number of grid cells with average DLI exceeding 9$\Phi$ in the month of March. 

The range of observed average DLI correspond to the DLI requirements of crops that belong to the very low light (< 5$\Phi$), low light (5--10 $\Phi$), and moderate light (10--20 $\Phi$) categories. Out of these categories, only crops grown under moderate light conditions are considered suitable for commercial production \citep{faustnd}. Thus, crops such as sweet pepper (\textit{Capsicum annuum}) and lettuce (\textit{Lactuca sativa}) belonging to the moderate-light categories can be grown at locations where average DLI exceeds 9$\Phi$ on these fa\c{c}ades \citep{songetal2018}.
  
Statistically significant and moderate to high values of $\rho$ (> 0.5) under different skytypes suggests that positive linear relationship exists between measured and simulated PAR, affirming the usability of 3D city models for this use case. High values of MAE and RMSE suggest that the simulated PAR deviate from the measured PAR. Thus, the simulated PAR was able to capture the trend followed by measured PAR at a sensor location but not its values. There are mainly three reasons for this. Firstly, the 3D model used for carrying out simulations was LOD1 model, excluding vegetation. The present model did not take into account the architectural elements of the fa\c{c}ades such as cantilevered louvres with perforated sunscreen above the windows, roof overhangs on level 16, among others and the presence of trees causing shadows at lower levels. As a result, simulations do not account for their effects on PAR at a location. Although higher detailed (i.e.\ LOD3) models can take care of fa\c{c}ade's architectural elements, past research suggests that these models are generally available for a small study area and their non-availability becomes a constraint when considering a larger study area.

\begin{figure}[pos = htbp]
	\centering
	\includegraphics[width=0.5\textwidth]{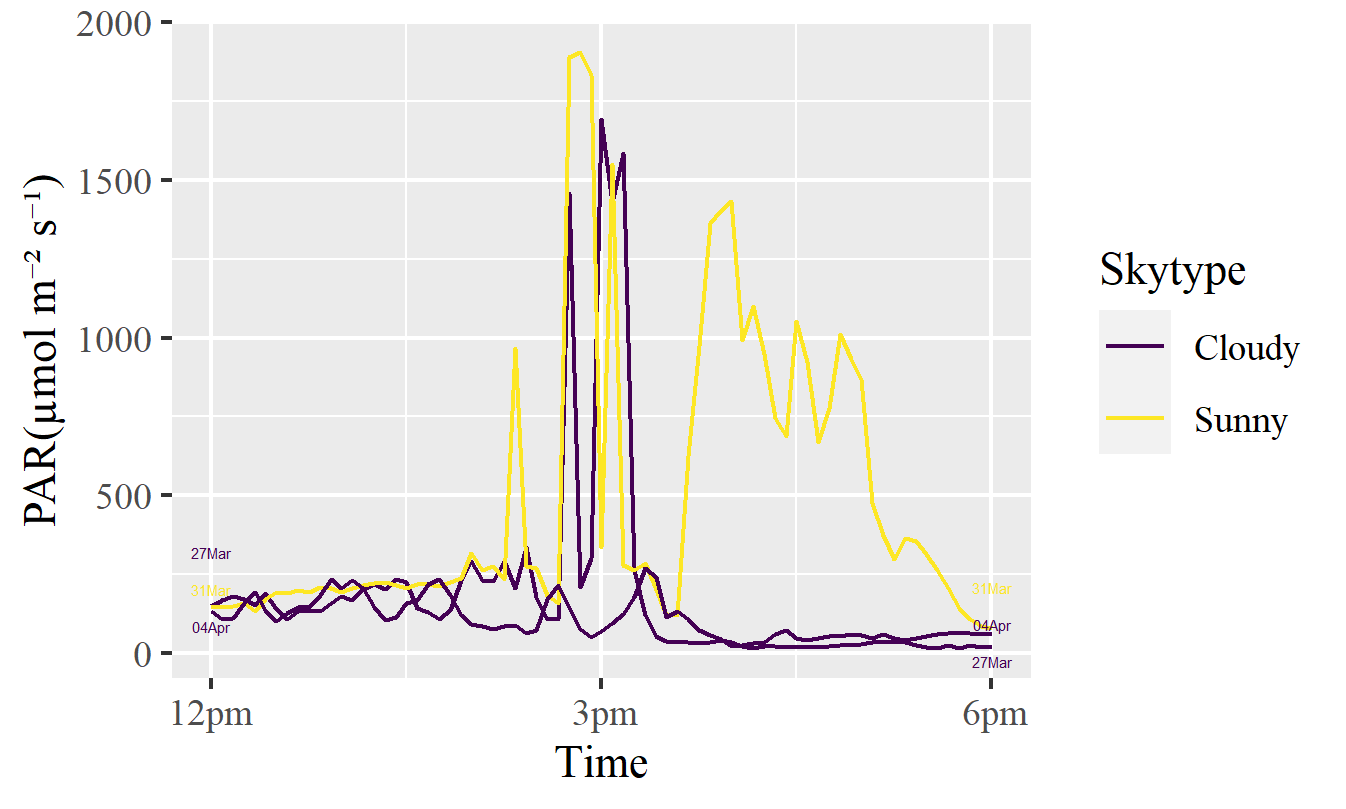}
	\caption{\label{fig:9acalib}Measured PAR at location 9a on sunny and cloudy days during period III.}
\end{figure}

Secondly, hourly simulated PAR generated at various sensor locations for comparison with measured PAR were based on the 24-hour weather forecasts and not the actual weather conditions. Cases have emerged where PAR measured through sensors on a cloudy day were found to be equivalent to those measured on a sunny day. For instance, as seen in Figure \ref{fig:9acalib}, PAR logged at location 9a around 3pm on 27 Mar (cloudy day) matched to those measured on 31 Mar (sunny day). PAR logged around 3pm on 27 Mar were very high as compared to those logged for another cloudy day (04 Apr) around the same time. Such differences in actual and forecasted weather conditions have resulted in exceptionally low $\rho$ and high MAE and RMSE for the cloudy sky conditions on fa\c{c}ade W. Relatively low $\rho$ and relatively high MAE and RMSE observed at sensor locations on fa\c{c}ade B and level 14 of fa\c{c}ade C for partly cloudy and cloudy sky conditions can also be attributed to the same reason. These outliers have significantly impacted the performance of otherwise fairly accurate simulations. While it may be possible to reduce this error source by using actual weather conditions, this shall go beyond the scope of this paper due to non-availability of a credible source of open data at this temporal scale and due to large variability of cloud cover between locations in Singapore \citep{tanismail2015}.
 
\begin{figure}[pos = htbp]
	\centering
	\includegraphics[width=0.5\textwidth]{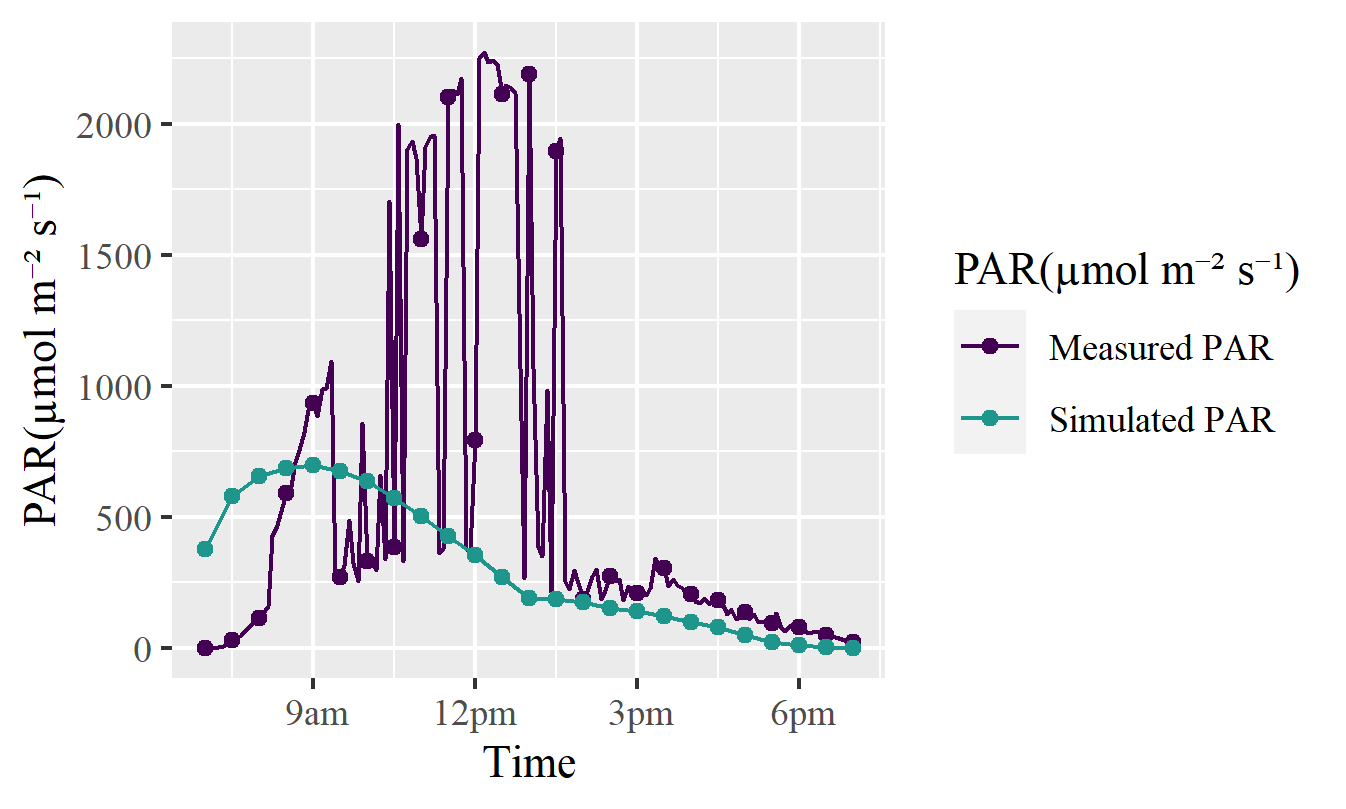}
	\caption{\label{fig:12B1calib}Measured and simulated PAR for location 12a on a sunny day (02 Mar), which has the highest discrepancy.}
\end{figure}

Lastly, the inability of these simulations to model the sharply contrasting periods of low and high PAR is another reason for deviation in measured and simulated PAR. To illustrate the same, Figure \ref{fig:12B1calib} shows measured and simulated PAR at location 12a on a sunny day (02 Mar) at five-minute and half-hourly intervals respectively with PAR at half-hourly intervals highlighted with points ($\bullet$). This particular location is free from any obstacles that may possibly affect measured PAR. As seen in the figure, this high variability in PAR is observed around noon when the sun is positioned right above and moving past fa\c{c}ade B. Consequently, such variable sunlight conditions \citep{smith2013} can be attributed to the parapet edges of the higher levels which have been captured by the PAR sensor also placed on the railing of the corridor's parapet (Figure \ref{fig:placementCorridor}). Same reasoning can also be applied to sensors placed at levels 8, 10, and 15. On the same level of fa\c{c}ade B, relatively better values of these metrics for location `b' (e.g. 12b) than location `a' (e.g. 12a) is due to the fact that these locations were shadowed by fa\c{c}ade C for some duration (e.g. 7am-1pm in Figure \ref{fig:shadowMaps}) when this fa\c{c}ade received direct sunlight. As a result, some of these variable sunlight episodes were not observed in measured PAR for these locations. Given these reasons, simulated PAR data compare fairly well with measured PAR.
 
The excessive irradiation captured by the PAR sensor over short durations around noon also contribute to the DLI at the sensor location. Consequently, these variable sunlight episodes may result in crop's reduced photosynthetic performance in the built environment \citep{tanismail2014, songetal2018} due to its mistaken selection based on inflated DLI. By not being able to model these episodes, simulations eliminate locations that receive higher DLI due to these bouts of highly variable intensity of sunlight. At the same time, this also results in underestimation of DLI at a location found suitable for farming. Thus, some caution may have to be exercised during crop selection. Only those crops having threshold DLI obtained through simulations and which are tolerant to these episodes would be suitable for such locations.

\begin{figure*}[pos = htbp]
	\centering
	\includegraphics[width=0.7\textwidth]{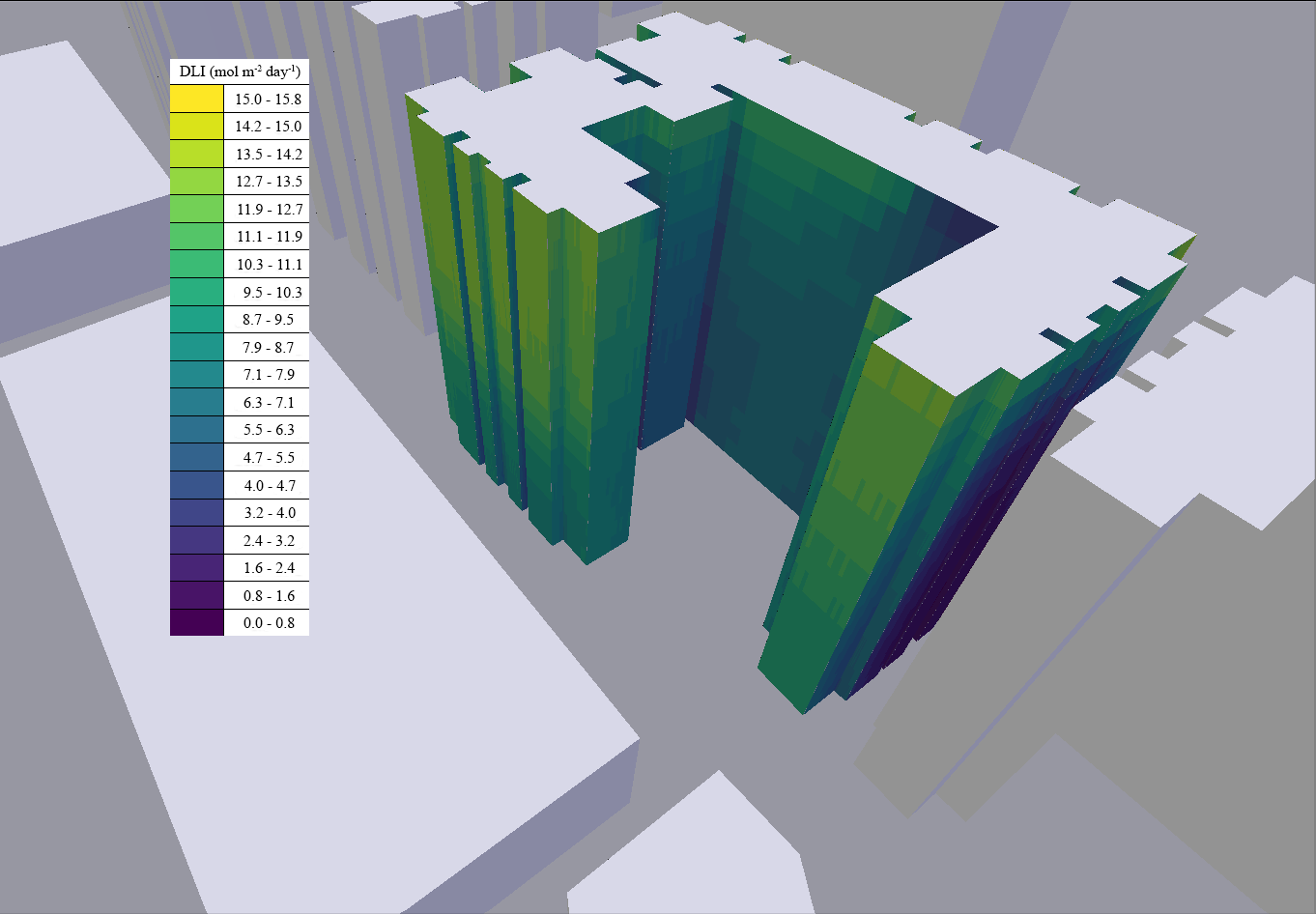}
	\caption{\label{fig:facadesDLI}Average DLI for March on fa\c{c}ades of the study area. These results are a critical insight for decision-making for high-rise urban farming and for maximizing the crop yield.}
\end{figure*}

Sunlight availability is often the limiting factor for crop growth when sufficient water and nutrients have already been supplied to the crops in well managed farming systems. Various studies have used PAR to predict potential productivity of crops with known leaf area index and light use efficiency \citep{zhaoetal2010,pulestonetal2017}. However, such predictions were only conducted for traditional large-scale farmlands. To do similar prediction for vertical spaces of buildings, it would be necessary to estimate PAR and DLI in buildings using 3D modeling.  The results in the previous section and the above discussion have not only corroborated the findings of the existing literature but have also contributed toward an enriched understanding of PAR and DLI at different micro-locations of the same building as well as in different sky conditions. Particularly, it highlighted how 3D city models can facilitate understanding of the sun's diurnal and annual motion in the study area and enable estimating the shadowing effects and DLI for locations that may not be easily accessible to PAR surveys (e.g. windows of dwelling units). Based on these estimations, crops such as lettuce and sweet pepper were found suitable for growing in the vertical spaces of the study area. However, the presence of various architectural elements on these vertical spaces diminish the possibilities of large scale commercial farming and may only favour subsistence farming. Thus, with 3D city models it is possible to assess the suitability of a micro-location in a building for farming; alleviating the need to conduct PAR survey and simultaneously showcasing their new application. Insights drawn from these analyses for rooftops, fa\c{c}ades, and other potential farming locations in buildings can be used by decision makers to assess farming potential at urban scale at a glance. For instance, they can map the locations of grid cells found suitable for farming (Figure \ref{fig:facadesDLI}) with corresponding locations in 3D modeling tools such as Blender and Google Earth (Figure \ref{fig:urbanForms}) and examine whether it is practically possible to do farming at those locations. This can then be further utilized to guide urban farmers in crop selection and appropriate cropping cycle based on DLI estimations along with best farming practices to maximize their agricultural produce. Urban farming companies can also benefit from such analyses while performing their on-site assessments and identifying potential urban farming sites. They would also be able to predict their yearly crop yield based on the DLI estimations at different locations of the building. The 3D modeling approach together with the predicted crop yield can also help local government to predict the level of self-sufficiency the city can achieve and to better plan their land use policies \citep{diehl2020}. Not to mention, while the present paper has only focused on a particular building in Singapore, the methodology employed to generate 3D city models and the analyses carried out herein are equally applicable to other high-rise buildings within and beyond Singapore.
     
\subsection{Limitations}
While the results and subsequent discussion suggest that 3D city models can support urban farming site identification in buildings and help deciding which crop to grow at which site, they suffer from some limitations. Firstly, EPW data used in DLI simulations in this paper correspond to the 1990s. As a result, the actual DLI in the present conditions may differ from the simulated DLI. However, significant variations in DLI at the scale of months/year, which have been used in the analysis, are not expected given the equatorial position of Singapore. Secondly, besides the limitation associated with ground reflections, this version of VI-Suite does not account for leap year, 2020 being one. While large variations in simulated PAR were not observed between consecutive days, it is hoped that this and other limitations will be rectified in its upcoming versions. Lastly, vegetation around buildings may play an important role in solar exposure assessment, which we did not have in our 3D city model.

\section{Conclusion}\label{sec:conclusion}

This paper investigated a new application of using 3D city models to identify urban farming sites in buildings and understand their potential for growing particular crops based on sunlight properties derived by simulations. It capitalized on the prior work relying on 3D city models to estimate the solar potential for assessing the suitability of installing photovoltaic panels on rooftops, and adapted it for a significantly different purpose and of a different nature with certain particularities (urban farming) and different locations (vertical spaces of buildings). Our work includes field measurements to verify the integrity of the simulations, which is a rarity in related work. The important points from this work are:
\begin{itemize}
    \item There is a large variation in the level of available sunlight within a building, requiring understanding the potential of different sites for urban farming at a micro-location scale.
    \item 3D city models can be conveniently used to support urban farming by identifying such sites in an approximate manner. They have an unparalleled advantage over doing field measurements when there are many more locations to evaluate and especially when scaling up the estimations at the precinct or urban scale to cover thousands of buildings, which is in practice impossible to carry out with field measurements.
    \item Such analyses can be conducted using block (LOD1) models obtained from open data, and the simulations can be run using open-source software, facilitating replication elsewhere and scalability to cover entire cities.
\end{itemize}

We believe that this novel use case has rich potential to be further researched, and there are several avenues for expanding this work. Quantifying the farming area and the projected crop yield in a building is the first one \citep{shaoetal2016}. Improving simulation accuracy by employing datasets including vegetation and using 3D models of higher detail such as architectural models which are becoming increasingly common \citep{Biljecki:2021vy}, integrating dynamic and indoor data which is a promising research direction in 3D city modelling~\citep{Kutzner:2020ij,Konde:2018ew}, and experimenting with different material types for ground surfaces offer another line of future work. Most importantly, growing crops at farming locations identified through simulations would be the real test of 3D city models.
It is hoped that these lines of research will show the path to accurately estimate the farming potential in buildings and provide thrust to undertake this activity at the urban scale.
As another possible future work direction, it is also foreseen that once the potential is assessed and the ensuing urban farming activities in buildings commence, 3D city models --- optionally coupled with additional data such as legal matters --- can be used also to manage them and serve as a registry for coordination purposes, for example, for organising the provision of subsidies and for issuing permits for farming in public buildings.
Finally, it would be interesting to investigate whether this use case can be combined with assessing the suitability for installing solar panels and energy simulations, recommending the optimal mix and arrangement of photovoltaic installations and agricultural crops in the same building, presenting a holistic solution for supporting green buildings and sustainable development.

\section*{Acknowledgements}
The authors thank the anonymous reviewers for their helpful comments. The authors appreciate the contact with Dr Ryan Southall (University of Brighton) for answering queries related to VI-Suite, Dr Siu-Kit Lau (National University of Singapore (NUS)) for providing a light meter during the preliminary analysis, and the mapping efforts of the OpenStreetMap community. The authors would also like to thank Dr Puay Yok Tan (NUS) and Dr Xiao Ping Song (NUS) for their help with the PAR surveys.
This research was conducted as part of the master's thesis of the lead author while he was on sabbatical leave from Union Bank of India. It is also part of the project Large-scale 3D Geospatial Data for Urban Analytics, which is supported by the National University of Singapore under the Start Up Grant R-295-000-171-133.

\appendix
\section{Appendix} \label{sec:appendix}

\begin{table}[cols=3,pos=h]
\caption{Calibration equations for the PAR sensors used in the survey.}
\label{tbl1}
\begin{tabular*}{\tblwidth}{@{} LLL @{} }
\toprule
PAR sensor ID & Calibration equation & R$^{2}$\\
\midrule
2258-3 & y = 0.9918x + 79.884 & 0.9034\\
2558 & y = 0.861x + 98.593 & 0.8953\\
2559 & y = 0.9802x - 28.555 & 0.9009\\
2896-5 & y = 0.952x - 18.332 & 0.9021\\
2897-8 & y = 0.9133x + 65.485 & 0.9012\\
2902-10 & y = 0.9448x - 46.742 & 0.8982\\
2904-9 & y = 0.9485x + 4.5319 & 0.9071\\
8982-7 & y = 1.0185x - 38.092 & 0.9004\\
8983-6 & y = 0.9928x + 7.4332 & 0.9042\\
8986-4 & y = 1.0154x - 24.251 &	0.9030\\
\bottomrule
\end{tabular*}
\end{table}

\begin{table}[cols=3,pos=h]
\caption{Parameter values for simulations in VI-Suite.}
\label{tbl2}
\begin{tabular*}{\tblwidth}{@{} LLL @{} }
\toprule
Analysis/Simulation type & Parameter name & Parameter value\\
\midrule
\multirow{2}{4em}{Sun path} & Suns & Single or Hourly \\ 
& Thickness & 0.15 \\ 
\midrule
\multirow{3}{4em}{Shadow map} & Animation & Static \\ 
& Result Point & Faces \\ 
& Offset & 0.01 \\
\midrule
\multirow{6}{4em}{Basic lighting} & Result Point & Faces \\ 
& Offset & 0.01 \\
& Program & Gensky \\
& Ground ref & 0.00 \\
& Turbidity & 2.75 \\
& Accuracy & Medium \\ 
\midrule
\multirow{4}{4em}{Climate Based Daylight Modelling} & Result Point & Faces \\ 
& Offset & 0.01 \\ 
& Type & Exposure \\
& Accuracy & Final \\
\bottomrule
\end{tabular*}
\end{table}

\begin{table*}[width = 0.9\textwidth, cols = 9]
\caption{Percent (\%) reduction in PAR under different sky conditions.}
\label{tbl3}
\begin{tabular*}{\tblwidth}{@{} LLLLLLLLL @{} }
\toprule
\multirow{3}{*}{Skytype} & \multirow{3}{*}{Time} & \multirow{3}{*}{Fa\c{c}ade} & \multirow{3}{*}{Level \#} & \multicolumn{2}{c}{Unshaded grid cell} & \multicolumn{2}{c}{Shaded grid cell} & \multirow{3}{*}{\% reduction}\\ \cline{5-8}
& & & & \multirow{2}{*}{\#} & PAR  & \multirow{2}{*}{\#} & PAR & \\
& & & & & ($\mu$mol m$^{-2}$ s$^{-1}$) & & ($\mu$mol m$^{-2}$ s$^{-1}$) & \\
\midrule
\multirow{5}{*}{Sunny (17 Mar)} & 10am & A & 13 & 10 & 568 & 9 & 259 & 54.40 \\ 
& 10am & B & 5 & 9 & 653 & 10 & 341 & 47.77 \\
& 10am & B & 10 & 12 & 570 & 13 & 256 &	55.08 \\
& 1pm & C & 5 & 5 & 434 & 4	& 240 &	44.70 \\
& 4pm & C & 13 & 11	& 606 & 10 & 84	& 86.13 \\
\midrule
\multirow{5}{*}{Partly Cloudy (09 Mar)} & 10am & A & 13 & 11 & 130 & 10 & 85 & 34.61 \\
& 10am & B & 5 & 10 & 151 & 11 & 98 & 35.09\\
& 10am & B & 10 & 12 & 145 & 13 & 90 & 37.93\\
& 1pm & C & 5 & 3 & 90 & 2 & 56 & 37.77\\
& 4pm &	C & 13 & 10 & 146 & 9 & 63 & 56.84\\
\bottomrule
\end{tabular*}
\end{table*}

\begin{table*}[width = 0.9\textwidth, cols = 5]
\caption{Monthly average DLI range and levels with average DLI above 9 mol m$^{-2}$ day$^{-1}$.}
\label{tbl4}
\begin{tabular*}{\tblwidth}{@{} LLLLL @{} }
\toprule
\multirow{3}{*}{Month} & \multirow{3}{*}{Fa\c{c}ade} & \multicolumn{2}{c}{Range} & Levels at which\\ \cline{3-4}
& & Minimum & Maximum & average DLI exceeds \\
& & (mol m$^{-2}$ day$^{-1}$) & (mol m$^{-2}$ day$^{-1}$) & 9 mol m$^{-2}$ day$^{-1}$ \\
\midrule
\multirow{4}{*}{March} & A & 5 & 12 & 14 and above \\
& B & 1 & 13 & all \\
& C	& 7	& 13 & all \\
& W	& 12 & 15 &	all \\
\midrule
\multirow{4}{*}{June} & A & 5 & 14 & 10 and above \\
& B & 1 & 10 & 15 and above \\
& C	& 4 & 10 & 14 and above \\
& W	& 11 & 13 & all \\
\midrule
\multirow{4}{*}{September} & A & 5 & 12 & 13 and above \\
& B	& 1	& 12 & 9 and above \\
& C & 6	& 12 & 9 and above \\
& W	& 12 & 14 & all \\
\midrule
\multirow{4}{*}{December} & A & 4 & 10 & 16 \\
& B	& 1	& 15 & all \\
& C	& 8	& 14 & 1, 2, 7 and above \\
& W	& 10 & 15 & all \\
\bottomrule
\end{tabular*}
\end{table*}


\printcredits

\newpage

\bibliographystyle{cas-model2-names}

\bibliography{cas-refs}


\end{document}